\documentclass[pra,amsmath,amsfonts,amssymb,notitlepage,showpacs]{revtex4-1}
\usepackage{bbm} 

\usepackage{bm} 
\usepackage{graphicx}
\usepackage{xcolor}

\begin{document}
\title{Stationary waves on nonlinear quantum graphs:\\
  General framework and canonical perturbation theory}

\author{Sven Gnutzmann} \affiliation{School of Mathematical Sciences,
  University of Nottingham, Nottingham NG7 2RD, UK} \author{Daniel
  Waltner} \affiliation{Fakult\"at f\"ur Physik, Universit\"at
  Duisburg-Essen, Lotharstra\ss e 1, 47048 Duisburg, Germany}

\begin{abstract}
  In this paper we present a general framework for solving
  the stationary 
  nonlinear Schr\"odinger equation (NLSE) on a network of one-dimensional
  wires modelled by a metric graph
  with suitable matching conditions at the vertices.
  A formal solution is given that expresses the wave function
  and its derivative at one end of an edge (wire) nonlinearly in
  terms of the values at the other end.
  For the cubic NLSE this nonlinear transfer operation 
  can be expressed explicitly in terms of Jacobi elliptic functions.
  Its application
  reduces the problem of solving the corresponding set
  of coupled ordinary nonlinear differential equations to a finite
  set of nonlinear algebraic equations. 
  For sufficiently small amplitudes we use canonical perturbation theory
  which makes it possible to extract the leading nonlinear corrections over large distances.
\end{abstract}


\maketitle

\section{Introduction}
\label{intro}

In a series of two papers we treat stationary solutions on nonlinear
quantum graphs and introduce an approach based on
canonical perturbation theory.  This is the first paper in the series where we deal
with the general theoretical framework. In the second paper \cite{GW2}
we will apply the framework to a set of basic graph structures.

Linear quantum graphs where the wave function 
obeys the linear Schr\"odinger equation on
the edges with suitable matching conditions  
have attracted a lot of attention in physics and mathematics in the past
(see \cite{Kottos,GS_Adv,Berkolaiko} and references therein). 
In quantum chaos they serve as
a paradigm model that makes it possible to analyze spectral fluctuations 
\cite{Kottos,Kottos1,GS_Adv,GA}, 
wave function statistics \cite{GKP}
and chaotic scattering \cite{Kottos2,Kottos3,Weidenmueller}.
More generally they are paradigm models for the effects of nontrivial
topologies on wave function propagation. Nonlinear quantum graphs 
replace the linear wave equations with a nonlinear wave equation and
have the potential of becoming a paradigm model for topological
effects in nonlinear wave propagation because they are sufficiently
simple to allow for comparatively straight forward numerical analysis
and analytical approaches while showing fundamentally nonlinear
effects (such as multistability or bifurcations).

Physically they can be considered as models for wave
propagation in optical networks and quasi-one-dimensional (cigar-like)
Bose-Einstein condensates \cite{Zhang}. 
In either optical systems or Bose-Einstein condensates 
nonlinear effects enter naturally. In the optical systems 
this is due to nonlinear media (Kerr effect) and in 
Bose-Einstein condensates it is due to the
boson-boson interaction. It is then required to add nonlinear terms
to the Schr\"odinger equation, which makes explicit analysis generally 
much harder. One is often restricted to either
numerical analysis (see, e.g.,~\cite{Paul}), diagrammatic approaches valid
for small nonlinearities (see, e.g.,~\cite{Hartung}) or to
one spatial dimension \cite{Rapedius,Leboeuf}. 

Adding nonlinear terms to the wave equation on a quantum graph
results in a nonlinear quantum graph. A numerical survey \cite{Gnutzm} 
showed the importance of nonlinear effects in stationary scattering from
a nonlinear graph even if the incoming waves have very low intensity.
As has been revealed later \cite{Gnutz1}, this is partly due to the presence 
of very narrow (so-called topological) resonances. 
Stationary solutions on nonlinear quantum graphs have been
discussed for some basic graph structures \cite{Lytel1}.
For a general nonlinearity proportional to $|\psi|^{2\nu}\psi$,
stationary states on a star graph were considered
\cite{Adami3,Adami6,Adami2}. The phase space structure was analyzed on a
three edge star graph \cite{Adami5} and the stability of the states was
studied \cite{Adami4}. 
On a tadpole graph bifurcations and stability
of stationary solutions have been analyzed \cite{tadpole1,tadpole2}.

Some time-dependent solutions have been considered. The propagation
of a soliton through a vertex in a star graph was analyzed in
\cite{Adami,Sobirov,Holmer,Uecker,Caudrelier}. Interacting Bose liquids in Y-junctions and
ring geometries \cite{Tokuno} and H-shaped potentials \cite{Viet}
or several differently connected branches of discrete nonlinear networks
\cite{Miros,Burioni,Stojanovic}
may also be considered as nonlinear quantum graphs. 
An experimental realization of
one-dimensional scattering in optical nonlinear media is reported in
\cite{Linzon}; the escape of solitons in \cite{Peccianti}.
A recent review by Noja \cite{NojaReview} summarizes nicely some of the more mathematical
approaches mentioned above.

Our first aim in this paper is to
reduce the coupled nonlinear differential equations to a finite set
of nonlinear algebraic equations.
Such a reduction requires the solution of the nonlinear transfer problem;
i.e., one needs to express the wave function and its derivative at one end 
of an edge in terms of the values at the other end.
We give a general formal solution to this problem. 
For the cubic nonlinearity these can be expressed
explicitly using Jacobi elliptic functions. In an extensive appendix
we give the complete set of
solutions in this case. This extends the known explicit stationary 
wave functions on a line or a circle \cite{Carr1,Carr2}. 
The second aim is to 
develop a perturbation theory that simplifies the formal transfer
solution such that analytical methods can be used to
find approximate wave functions on a graph.

In Section~\ref{nlgraphs} we give the general framework and discuss
general properties of the stationary solutions:
We define nonlinear quantum graphs, describe how to obtain local solutions 
on the edges
and  explain how to reduce the general problem of finding
stationary solutions to a finite set of nonlinear algebraic
equations. 
In Section~\ref{canon_pert} 
we introduce a perturbative treatment of the nonlinearity in the framework of
canonical perturbation theory. 
Two extensive Appendixes \ref{sec:analytic_solutions} and
  \ref{angularfrequencies} contain detailed explicit solutions for the
  cubic NLSE for reference.

\section{Nonlinear Quantum Graphs}
\label{nlgraphs}

\subsection{General Setting}
\label{sec:setting}

\begin{figure}[h]
  \begin{center}
    \includegraphics[width=0.6\textwidth]{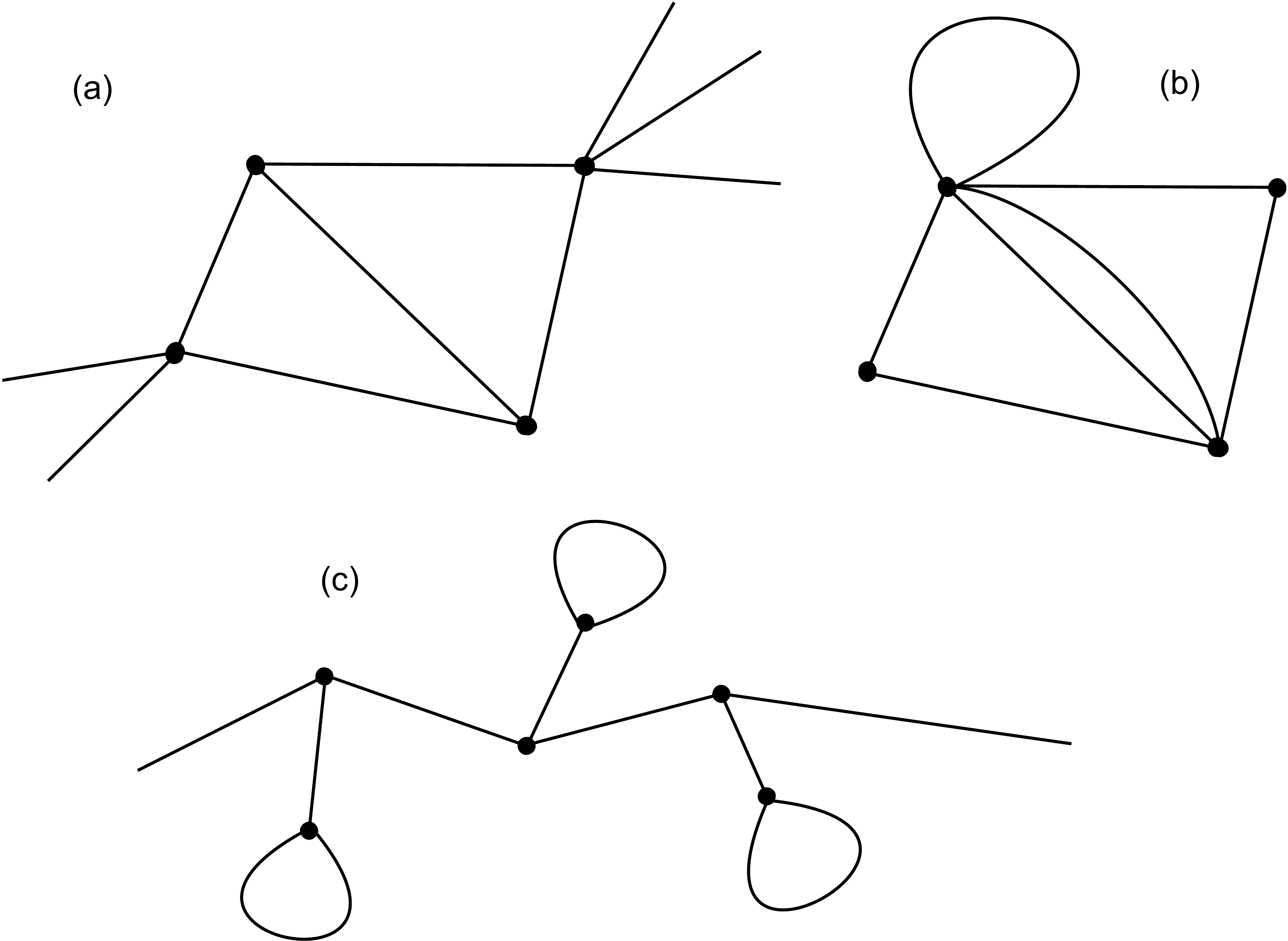}
    \caption{Examples of network structures: a finite open graph
      (a); a finite closed graphs (b). If the
      structure of the graph (c) is continued one obtains an
      infinite graph. \label{fig:graph}}
  \end{center}
\end{figure}

We consider a general graph $\mathcal{G}(\mathcal{V},\mathcal{E})$
where $\mathcal{V}$ is a set of vertices and $\mathcal{E}$ a set of
edges.  In standard graph theory each edge $e \in \mathcal{E}$
connects two vertices $v_1, v_2 \in \mathcal{V}$.  If $v_1=v_2$, the
edge is called a loop. Two different vertices $v_1\neq v_2$ are
called adjacent or connected if there is an edge that connects
them. In this case we also say that the edge is connected or adjacent
to the vertices $v_1$ and $v_2$.  
In the present context it is
useful to generalize the notion of a graph slightly and allow for
\emph{semi-infinite} edges that are only connected to one vertex
(formally one may think of this as a standard graph with one vertex
`at infinity' that collects all the loose ends).  We call the
semi-infinite edges \emph{leads} and all edges that connect two
vertices (including loops) \emph{bonds}.  The corresponding sets of
leads and bonds are denoted by $\mathcal{L}$ and $\mathcal{B}$ and we
have $\mathcal{E}= \mathcal{L} \cup \mathcal{B}$ and $\mathcal{L} \cap
\mathcal{B}= \varnothing$.  We only consider \emph{finite} graphs
where the number of vertices $V=\left|\mathcal{V}\right|$ 
and edges $E=\left|\mathcal{E}\right|$ are both finite.  The numbers
of leads 
$L=\left|\mathcal{L}\right|$ of and bonds $B=\left|\mathcal{B}\right|$
are then also finite and $E=B+L$.  If a graph has no leads
$L=0$ we call it a \emph{closed graph}, if it has at least one lead $L
\ge 1$ then we call it an \emph{open graph}.  Figure \ref{fig:graph}
shows examples of open and a closed graphs.

In a quantum graph each edge models a waveguide or string with
waves propagating along them. This is realized by adding a
metric and a well-defined wave equation on the graph.  In a metric
graph each edge has a length $\ell_e>0$, and a coordinate $x_{e} \in
[0,\ell_e]$. For any bond $b \in \mathcal{B}$ the length is finite $\ell_b < \infty$
and $x_b=0$ and $x_b= \ell_b$ correspond to the endpoints of the
edge.   For each lead $l \in
\mathcal{L}$ the length is infinite $\ell_l= \infty$, and $x_\ell=0$
at the vertex attached to the lead. This structure defines the
distance between two points anywhere on the graph in an obvious way as
the length of the shortest connected path through the graph that
connects the two points \cite{footnote1}.

We consider a scalar complex wave function on the metric graph
which is differentiable with respect to $t$ and with respect to $x$ on
the edges.  It is written as a collection,
\begin{equation}
  \Psi(x;t)=\{ \psi_e(x_e;t) \}_{e\in \mathcal{E}},
\end{equation}
where $\psi_e(x_e;t)$ is the wave functions on the edge $e$ at time
$t$.  The wave function on edge $e$ satisfies the nonlinear
Schr\"odinger equation (NLSE)
\begin{equation}
  i \partial_t \psi_e =-  \partial^2_{x_e} \psi_e 
  + g_e |\psi_e|^{2\nu} \psi_e   \ .
  \label{NLSE_Time_dependent}
\end{equation}
Here, $\nu>0$ characterizes the power of the nonlinearity and $g_e$ is
the real nonlinear coupling parameter which we assume constant and
finite $|g_e|<\infty$ on each edge.  The nonlinear coupling term is
called repulsive for $g_e>0$ and attractive for $g_e<0$.  Without loss
of generality we use units where Planck's constant and the mass take
values $\hbar=2m=1$ everywhere in this paper. The \emph{cubic} NLSE
that is relevant for Bose-Einstein condensates or optical media
is obtained when $\nu=1$ -- in this case the
nonlinear term $g |\psi|^2 \psi$ is cubic in $\psi$.  
The quintic case $\nu=2$ with nonlinear term $g|\psi|^4\psi$ also finds some
applications \cite{Berge}.

Matching conditions at the vertices need to
be added to have a well-posed propagation of an initial wave function
$\Psi_0(x)$. For linear quantum graphs the most general matching
conditions that result in a self-adjoint problem have been given in
\cite{KostrykinSchrader} (see also \cite{Berkolaiko}).  All of these
remain mathematically and physically sound in the nonlinear setting
(and may be generalized by allowing nonlinear matching conditions).
Here we focus on the so-called $\delta$-type (or Robin) conditions.
For a given vertex $v$ let us assume (without loss of generality) that
$x_e=0$ corresponds to the endpoint at $v$ for all edges $e$ adjacent
to $v$.  Then $\delta$-type matching conditions at $v$ are defined by
two conditions:
\begin{enumerate}
\item[i.] The wave function is continuous through the vertex,
  \begin{equation}
    \psi_e(0;t) = \psi_{e'}(0;t) \equiv \psi_0,
    \label{matching_continuity}
  \end{equation}
  for all pairs of edges $e, e' \in \mathcal{E}(v)$ and all times $t$.
  By definition, $\mathcal{E}(v)$ is
  the set of all edges connected to
  the vertex $v$.
\item[ii.] The sum of outward derivatives of the wave function on the
  adjacent edges is proportional to the value of the wave function on
  the vertex
  \begin{equation}
    \sum_{e \in \mathcal{E}(v)}\ \partial_{x_e} \psi_e (0;t) 
    = \lambda \psi_0\ .
    \label{matching_flux}
  \end{equation}
  Here $\lambda$ is a real parameter, the \emph{vertex
  potential}. For a vertex of valency two this condition is equivalent
  to a $\delta$-potential of strength $\lambda$ on an interval (the
  position of the $\delta$-potential marking the position of the
  vertex).  For $\lambda<0$ we call the vertex attractive and for
  $\lambda>0$ repulsive.  In most applications that we discuss
  later we choose $\lambda =0$. Then the matching conditions are
  also known as standard (aka free,  Kirchhoff or Neumann) matching
  conditions \cite{footnote2}.  
\end{enumerate}
In the linear setting ($g=0$) these conditions lead to a self-adjoined
extension of the metric Laplacian on the graph (i.e., a well-defined
Schr\"odinger operator).  The first condition is physically reasonable
and the second ensures that the $L^2$ norm of the wave function
\begin{equation}
  \left\| \Psi(x;t) \right\|^2= \sum_{e \in E} \int_0^{\ell_e} 
  |\psi_e(x_e;t)|^2 dx_e   
\end{equation}
is conserved. Physically, the $L^2$ norm corresponds to the number of
particles (number of atoms in a Bose-Einstein condensate, number of
photons or total intensity in optics).  If $g_e=0$ for all edges $e$,
Eq.~\eqref{NLSE_Time_dependent} becomes the (free) Schr\"odinger
equation on a metric graph and the model becomes a quantum graph. For
quantum graphs wave propagation is described by linear 
differential equations. If $g_e \neq 0$ on some edge the
differential equations are non-linear and we call the model \emph{nonlinear
quantum graph}. Note that a nonlinear quantum graph can be used as a
model of either a quantum mechanical system (Bose-Einstein condensate)
or a purely classical wave system (electromagnetic
waves in optical fibres).

Generally, one is interested in the time-dependent dynamics of
an (square integrable and sufficiently smooth) initial wave function
$\Psi_0(x)=\{ \psi_{e,0}(x_e)\}_{e\in\mathcal{E}}$. For an infinite line
this problem is formally solved by the so-called inverse scattering
method \cite{Zakharov}, which is practical only for soliton-like
solutions.  For a half line or an interval with appropriate boundary
conditions (e.g. Dirichlet) the problem of finding any time-dependent
solutions is highly nontrivial. The generalization of the method 
to star graphs has recently been discussed in \cite{Caudrelier}. 
In this paper we focus on
stationary solutions of the form
\begin{equation}
  \Psi(x;t)= e^{- i  \mu t}\Phi(x)  \quad  \Rightarrow \quad
  \psi_e(x_e;t)= e^{- i  \mu  t} \phi_e(x_e)\ .
\end{equation}
The function $\Phi(x)=\{\phi_e(x_e)\}_{e=1}^E$ is then a collection of
solutions of the stationary NLSE
\begin{equation}
  -  \frac{ d^2 \phi_e}{dx_e^2} + g_e |\phi_e|^{2\nu} \phi_e 
  =  \mu \phi_e
  \label{NLSE_stationary}
\end{equation}
on each edge with the matching conditions \eqref{matching_continuity}
and \eqref{matching_flux} applied to $\Phi(x)$. We refer to the
parameter $\mu$ as the chemical potential (in accordance with the
physics literature on Bose-Einstein condensation).

\subsection{Formal Local Solutions on a Given Edge}
\label{sec:local_solutions}

Before discussing stationary solutions for a complete graph let us
first consider a single fixed edge $e$. We will
suppress the index $e$ until we come back to the discussion of the
full graph.  The length of the edge is $\ell$ and we assume that the
wave function and its derivative are given at $x=0$. Our aim is to
find the nonlinear transfer operator that expresses 
the wave function and its derivative at
$x=\ell$ in terms of their values at $x=0$. 
We will show that 
this is formally equivalent to the solution of an initial value
problem for a central force dynamics of a two-dimensional mass
point in the plane with a central
potential where $x$ takes the formal role of a time. 
The latter being integrable,
it is straight-forward to write a
formal solution using textbook methods of analytical mechanics.
Let us here summarize this approach and set
\begin{equation}
  \phi(x)= r(x) e^{i \eta(x)}
\end{equation}
with real amplitude $r(x)\ge 0$ and real phase $\eta(x)$.  
The NLSE is then expressed as  
two coupled real ordinary differential equations,
\begin{equation}
  \frac{d^2 r}{dx^2}= r\frac{d\eta }{dx}^2 
  +g r^{2\nu+1} - \mu r\quad \text{and} \quad
  \frac{d}{dx}\left[\frac{d\eta}{dx} r^2 \right]=0\ .
\end{equation}
If $x$ is formally considered a time
these equations  are the Euler-Lagrange
equations for a point particle in the plane in polar coordinates with
a central potential
\begin{equation}
  V(r)=  \frac{\mu}{2} r^2 - \frac{g}{2\nu+2} r^{2\nu+2}\ .
\end{equation}
The angular momentum
\begin{align}
  p_\eta = & r^2 \frac{d \eta}{dx} = \mathrm{Im}\ \phi^* \frac{d \phi}{dx}
             \label{eq:angular_momentum}
  \\
  \intertext{and the Hamiltonian energy}
  H= & \frac{1}{2} \frac{dr}{dx}^2 +\frac{p_\eta^2}{2 r^2} + V(r)
       \label{eq:energy}
\end{align}
are the two well-known constants of motion.  Note that in terms of the original NLSE
$p_\eta$ is the
intensity (or probability) flow. 
As is well known, the radial motion then reduces effectively to a
mass point in the effective potential
\begin{equation}
  V_{\mathrm{eff}}(p_\eta, r)= \frac{p_\eta^2}{2 r^2}+ V(r).
\end{equation}
Let us denote the solutions of the dynamical system with initial values
\begin{equation}
  \phi(0)=r_0
  e^{i \eta_0} \quad \text{and} 
  \quad
  \frac{d\phi}{dx}(0)= \left(\sigma
    \sqrt{2(H-V_{\mathrm{eff}}(r_0))} + i \frac{p_\eta}{r_0} \right)
  e^{i \eta_0}  
\end{equation}
(with $\sigma=\pm 1$)  as
\begin{equation}
  r(x)=R_{g,\mu}(x;r_0, p_\eta, H, \sigma)\quad \text{and}\quad
  \eta(x)=\eta_0 
  + \vartheta_{g,\mu}(x;r_0, p_\eta,H, \sigma)
\end{equation}
where the two functions $R_{g,\mu}(x;r_0, p_\eta, H, \sigma)>0$ and
$\vartheta_{g,\mu}(x;r_0, p_\eta, H, \sigma)$ are implicitly
defined through the two integrals
\begin{subequations}
  \begin{align}
    x= & \sigma \int_{r_0}^{R_{g,\mu}(x;r_0, p_\eta, H, \sigma)}
         \left(2(H-V_{\mathrm{eff}}(p_{\eta},r)\right)^{-1/2} dr
         \label{eq:Rg}
    \\
    \vartheta_{g,\mu}( x;r_0, p_\eta, H, \sigma )= & p_\eta
                                                     \int_0^x R_{g,\mu}(x';r_0, p_\eta, H, \sigma)^{-2} dx'\ .
                                                     \label{eq:thetag}
  \end{align}
\end{subequations}
We often just write $R_{g,\mu}(x)$ and $\vartheta_{g,\mu}(x)$ if
the values of the other parameters are clear from context.  
On the level of the NLSE these two functions implicitly define the nonlinear transfer operator
by evaluating them and their derivatives at $x=\ell$.
Note that
the integral \eqref{eq:Rg} defines $R_{g,\mu}(x;r_0, p_\eta, H,
\sigma)$ if $x$ is sufficiently small; for bounded solutions this
can be extended to arbitrary large values of $x$.  The sign $ \sigma=
\pm 1$ is positive (negative) if $R_{g,\mu}(x;r_0,
p_\eta, H, \sigma)$ is increasing (decreasing) as a
function of $x$ at $x=0$.\\
If the nonlinear coupling constant vanishes ($g=0$) the explicit
expressions for $R_{0,\mu}(x)$ and $\vartheta_{0,\mu}(x)$ can be
obtained from the known (local) solutions of the linear Schr\"odinger
equation
\begin{align}
  R_{0,\mu}(x) e^{i\vartheta_{0,\mu}(x)}= &
  \begin{cases}
    r_0\cos(kx) + \frac{\sigma \sqrt{2Hr_0^2-p_\eta^2-k^2 r_0^4} +i
      p_\eta }{kr_0}\sin(kx)
    & \text{if $\mu=k^2>0$}\\
    r_0 + \frac{\sigma \sqrt{2Hr_0^2-p_\eta^2}
      +i p_\eta }{r_0}x   & \text{if $\mu=0$}\\
    r_0\cosh (kx) + \frac{\sigma \sqrt{2Hr_0^2-p_\eta^2+k^2 r_0^4}
      +i p_\eta }{kr_0}\sinh(kx) & \text{if $\mu=-k^2<0$.}
  \end{cases}
  \label{eq:linsol}
\end{align}
The problem of finding solutions for arbitrary values of the chemical
potential $\mu \neq 0$ and nonlinear coupling constants $g \neq 0$ can
be reduced to a few standard solutions due to the scaling laws
\begin{subequations}
  \begin{align}
    R_{g,\mu}(x;r_0, p_\eta, H, \sigma)=& R_{\frac{g}{k^2},\pm
      1}\left(kx;r_0, \frac{p_\eta}{k}, \frac{H}{k^2},
      \sigma\right)
    \\
    \vartheta_{g,\mu}(x;r_0, p_\eta, H, \sigma)=&
    \vartheta_{\frac{g}{k^2},\pm 1}\left(kx;r_0, \frac{p_\eta}{k},
      \frac{H}{k^2}, \sigma \right)
  \end{align}
  \label{eq:scal1}
\end{subequations}
where $k= \sqrt{|\mu|}>0$ and
\begin{subequations}
  \begin{align}
    R_{g,\mu} (x;r_0, p_\eta, H, \sigma)= & \alpha R_{\pm 1,
      \mu}\left(x; \frac{r_0}{\alpha},
      \frac{p_{\eta}}{\alpha^2},\frac{H}{\alpha^2},\sigma\right)
    \\
    \vartheta_{g,\mu}(x;r_0, p_\eta, H, \sigma)= & \vartheta_{\pm
      1, \mu}\left(x; \frac{r_0}{\alpha},
      \frac{p_{\eta}}{\alpha^2},\frac{H}{\alpha^2},\sigma\right)
  \end{align}
  \label{eq:scal2}
\end{subequations}%
where $\alpha= \left| g  \right|^{-\frac{1}{2\nu}}$.\\
For a given exponent $\nu>0$ it suffices to consider the cases $g=
\pm 1$ and $\mu=\pm 1$ in order to get all local solutions for
arbitrary values $g$ and $\mu$ -- the case $\mu= 0$ is included by
taking the limits
\begin{subequations}
  \begin{align}
    R_{g,0}(x;r_0, p_\eta, H, \sigma)=& \lim_{k\to 0}
    R_{g,1}(kx;r_0, p_\eta/k, H/k^2, \sigma)
    \\
    \vartheta_{g,0}(x;r_0, p_\eta, H, \sigma)= & \lim_{k\to 0}
    \vartheta_{g,1}(kx;r_0, p_\eta/k, H/k^2, \sigma)
  \end{align}
\end{subequations}
and the limit $g\to 0$ needs to be consistent with \eqref{eq:linsol}.
\\
For the cubic NLSE  the functions
$R_{\pm 1,\pm 1}(x;r_0, p_\eta, H, \sigma)$ and $\vartheta_{\pm 1,
  \pm 1}( x;r_0, p_\eta, H, \sigma )$ can be expressed explicitly
in terms of Jacobi elliptic functions (see Appendix~\ref{sec:analytic_solutions}).  
For general $\nu$ we could not express the
integrals \eqref{eq:Rg} and \eqref{eq:thetag} in terms of any known
special functions. The qualitative behavior of
these solutions follows straight forwardly from the form of
the effective potential $V_{\mathrm{eff}}(r)$. While this is all 
well known, it is useful in the present context to summarize the
various cases. We do this in the rest of this section, adding some 
remarks related to their use in nonlinear quantum graphs.

\subsubsection{The repulsive case $g>0$}
\label{sec:repulsive}

\begin{figure}[h]
  \begin{center}
    \includegraphics[width=0.45\textwidth]{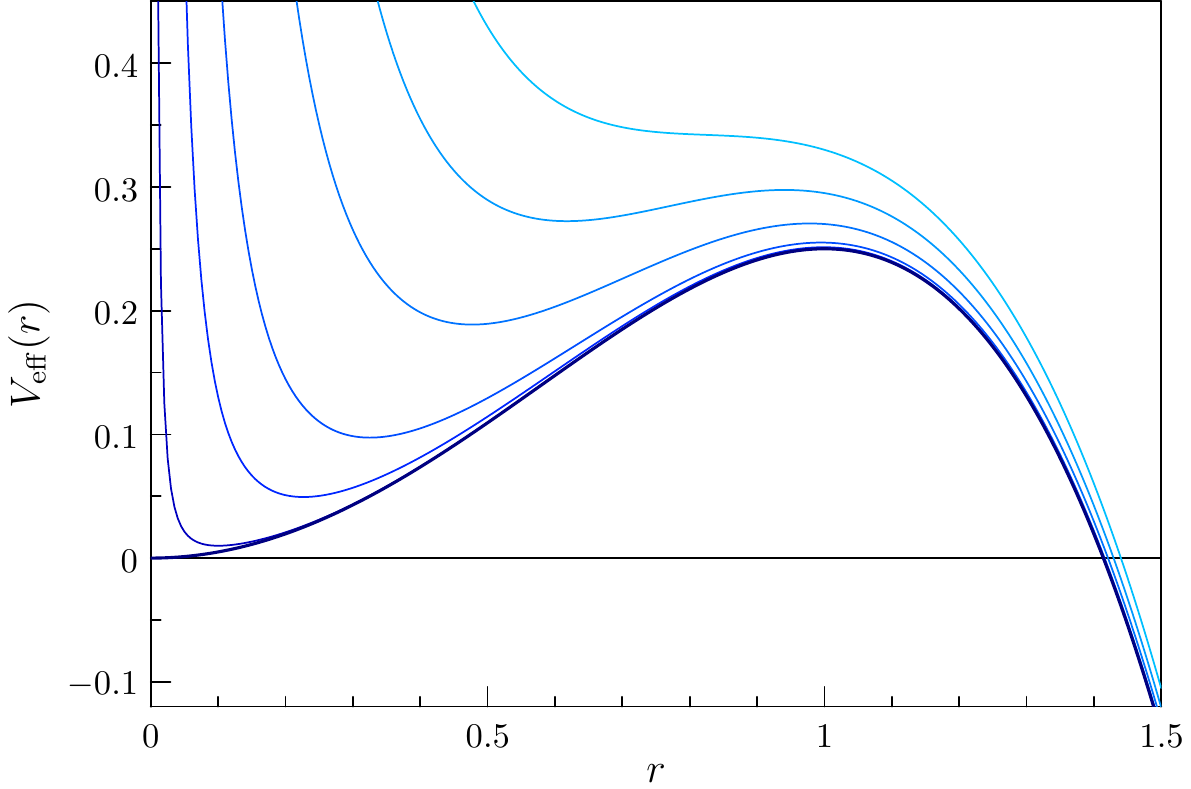}\hfill
    \includegraphics[width=0.45\textwidth]{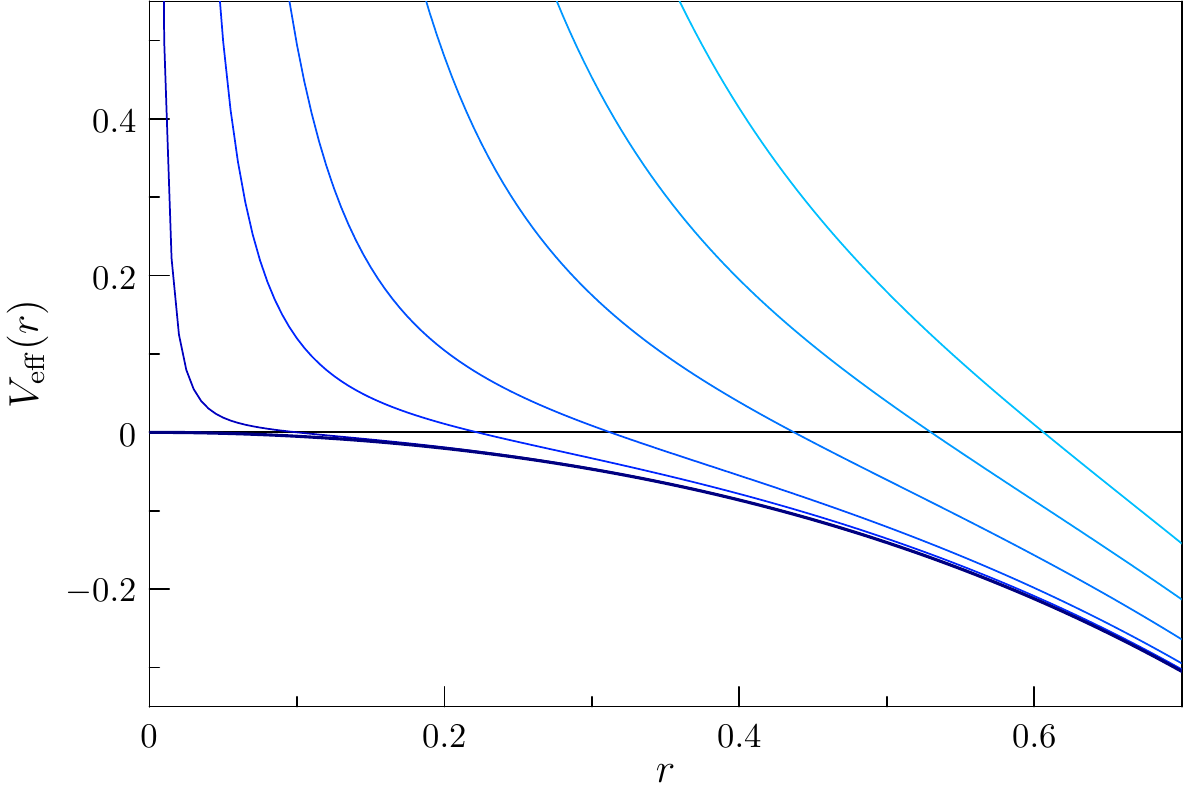}
    \caption{
      (Color Online) The effective potential $V_{\mathrm{eff}}(r)$  
      for the radial motion in the 
      auxiliary dynamical system
      with repulsive nonlinear coupling $g=1$ and $\nu=1$. 
      In the left graph the
      chemical potential is $\mu=1$; 
      in the right graph $\mu=-1$.
      The curves in each graph correspond to 
      different values of the
      flow (angular momentum of auxiliary dynamics) 
      $p_\eta=(0,0.01,0.05,0.1,0.2,0.3,0.4)$.\\
      Graphs for different values of the exponent 
      $\nu$ look qualitatively similar.
      \label{fig:Veff_repulsive}
    }
  \end{center}
\end{figure}

\begin{figure}[h]
  \begin{center}
    \includegraphics[width=0.45\textwidth]{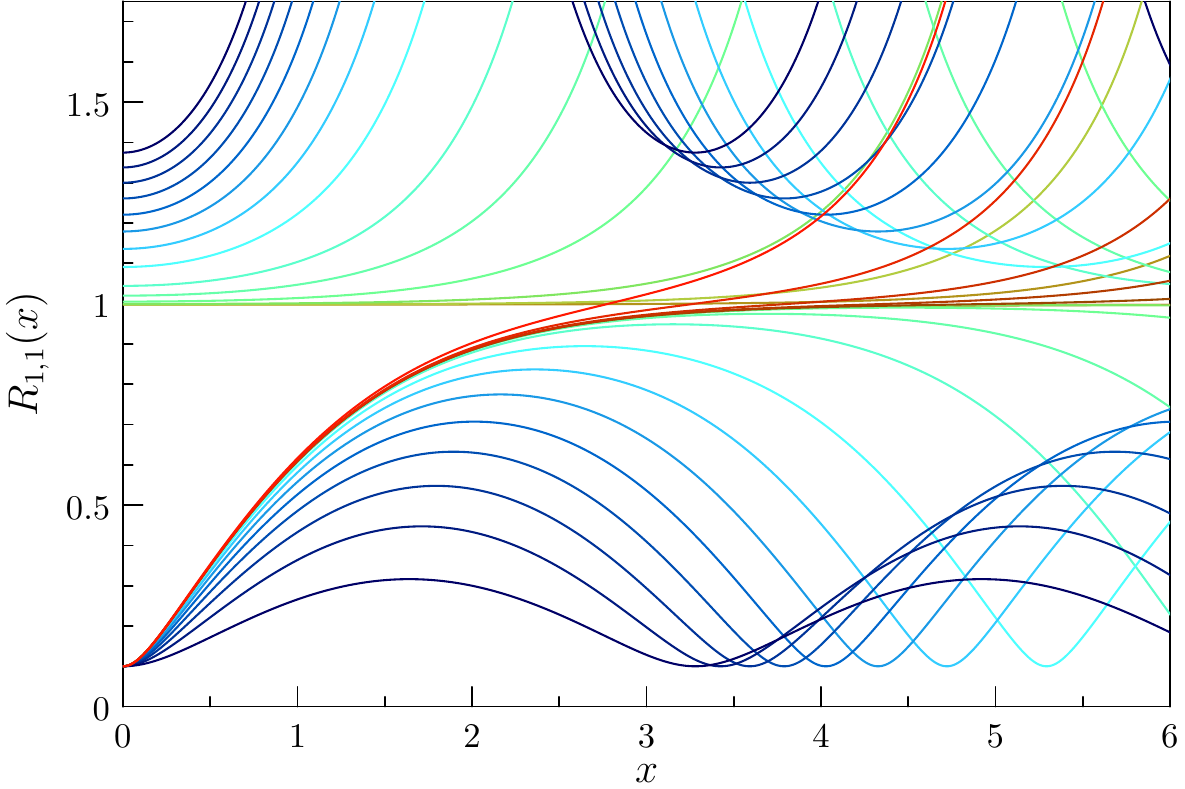}\hfill
    \includegraphics[width=0.45\textwidth]{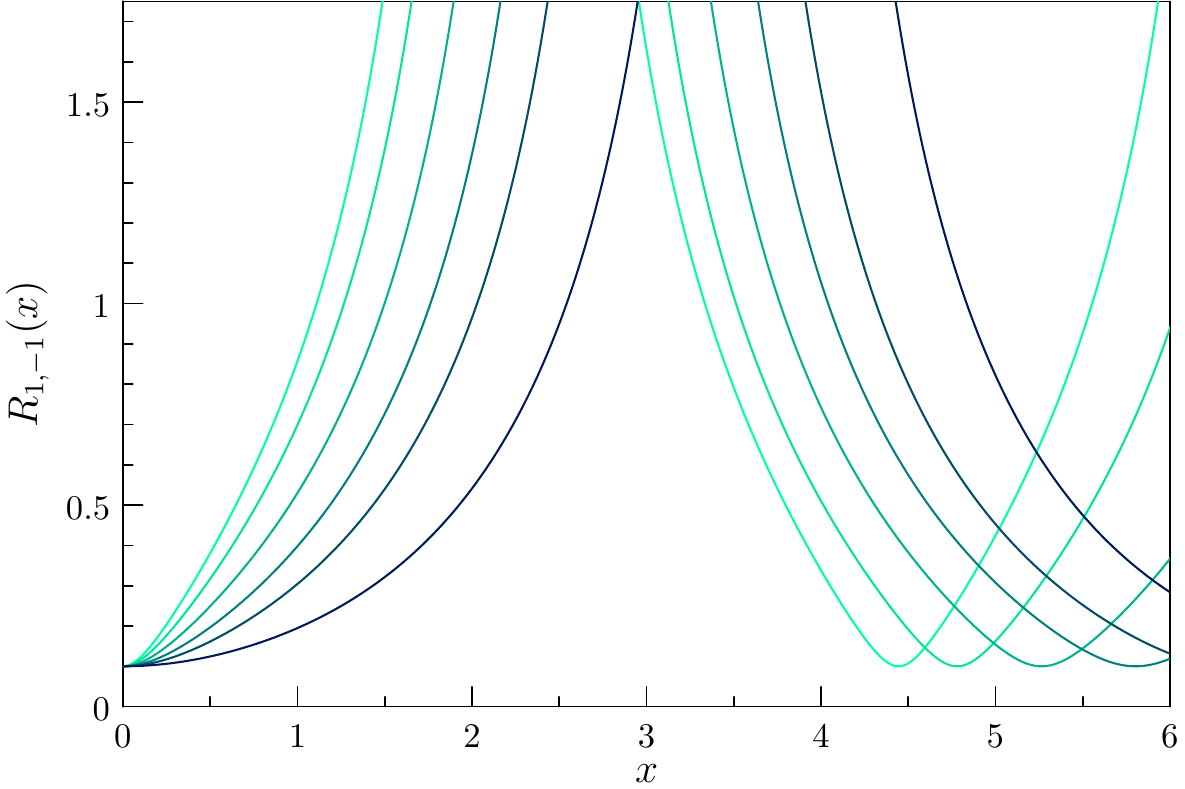}
    \caption{(Color Online) The amplitude functions $R_{1,1}(x)$ (left) and
      $R_{1,-1}(x)$ (right) for
      local solutions of the NLSE
      with repulsive interaction. 
      The parameters $H$, $p_\eta$ have been varied and $r_0$ 
      such that minimal amplitude is obtained for $x=0$.
      For $R_{1,1}(x)$ bounded and unbounded solutions are shown.
      The plotted functions are
      for $\nu=1$ ($\nu \neq 1$ leads to qualitatively similar
      solutions).
      \label{fig:R_repulsive}
    }
  \end{center}
\end{figure}

It is sufficient to consider $g=1$ and $\mu= \pm 1$.  
The solutions $R_{1, \pm
  1}\left(x; r_0, p_{\eta}, H,\sigma\right)$ and $\vartheta_{1, \pm
  1}\left(x; r_0, p_{\eta}, H,\sigma\right)$ depend mainly on the
two parameters $H$ and $p_\eta$ -- i.e. the
Hamiltonian energy and the angular momentum in the auxiliary
central potential dynamics in the plane.  The dependence on $r_0$ and
$\sigma$ is just a matter of shifting the origin $x \mapsto x-x_0$,
i.e., translating a solution.  For the qualitative discussion we
consider the effective potential
\begin{equation}
  V_{\mathrm{eff}}(r)= 
  \frac{p_\eta^2}{2 r^2} \pm \frac{r^2}{2} - \frac{1}{2\nu+2}
  r^{2\nu + 2}
\end{equation}
of the radial motion, where the sign is chosen positive (negative) for
$\mu=1$ ($\mu=-1$).  Figure \ref{fig:Veff_repulsive} shows the
effective potential for various values of the angular momentum
$p_\eta$.  For positive chemical potential $\mu>0$ we see the
following different types of solutions.
\begin{itemize}
\item[i.] For $|p_\eta| < p_{\mathrm{crit}}\equiv
  \left(\frac{\nu}{\nu+2}\right)^{1/2} \left( \frac{2}{\nu+2}
  \right)^{1/\nu}$ the effective potential has a local minimum and a
  local maximum at energies $H=\mathcal{E}_\mathrm{min}(p_\eta)$
  and $H=\mathcal{E}_\mathrm{max}(p_\eta)$ (see Figure
  \ref{fig:Veff_repulsive}).  For energies between these two energies
  $\mathcal{E}_\mathrm{min}\le H \le \mathcal{E}_\mathrm{max}$, there
  is a bounded solution where $r_1 \le R_{1,1}(x) \le r_2$
  and an unbounded solution
  $R_{1,1}(x)> r_3$ (see Figure \ref{fig:Veff_repulsive}).\\
  For the bounded solutions the amplitude is a periodic function
  $R_{1,1}(x)=R_{1,1}(x+\Lambda_{1,1})$ with period
  \begin{equation}
    \Lambda_{1,1}(H,p_\eta)= 
    2 \int_{r_1}^{r_2}
    \left(2H -\frac{p_\eta^2}{r^2} 
      - r^2 + \frac{1}{\nu+1} r^{2\nu+2}
    \right)^{-1/2} dr\ .
  \end{equation}
  This period tends to infinity when $H \to
  \mathcal{E}_\mathrm{max}(p_\eta)$ from below. In that case the
  amplitude $R_{1,1}(x)$ tends to the constant value
  $r_{\mathrm{max}}(p_\eta)$ for $x \to \pm \infty$ and has a
  single minimum at a finite $x$ value. Because of the corresponding
  dip in the amplitude such solutions are known as dark stationary
  solitons \cite{Ablowitz} though this name is sometimes reserved to
  the case $p_\eta=0$ where the intensity vanishes at one point.
  Figure \ref{fig:R_repulsive} shows plots for the amplitude
  $R_{1,1}(x)^2$ for various parameters including
  dark solitons.\\
  The phase function $\vartheta_{1,1}(x)$ is an increasing
  (decreasing) function if $p_\eta>0$ ($p_\eta<0$). In general,
  $\Delta \vartheta= \vartheta_{1,1}(x+\Lambda_{1,1}) -
  \vartheta_{1,1}(x)$ is not a rational multiple of $\pi$, so the
  corresponding wave functions $\phi(x)=R_{1,1}(x)
  e^{i\vartheta_{1,1}(x)}$ are, in general, not periodic functions of
  $x$. The wave function $\phi(x)=R_{1,1}(x) e^{i\vartheta_{1,1}(x)}$
  is real if and only if $p_\eta=0$. The phase does not change as
  long as the amplitude is positive. When $R_{1,1}(x)=0$ the wave
  function has a nodal point and changes its sign;
  i.e., $\vartheta_{1,1}(x)$ changes by $\pi$.  Indeed, it is clear from
  the form of $V_\mathrm{eff}(r)$ that $p_\eta=0$ if $\phi(x)$ has
  nodal points.  It can be shown explicitly that expression
  \eqref{eq:thetag} for the phase shows a discontinuous jump by $\pm
  \pi$ in the limit $p_\eta \to 0$. 
\item[ii.] For $|p_\eta < p_{\mathrm{crit}}|$ and either $H >
  \mathcal{E}_\mathrm{max}$ or $H < \mathcal{E}_\mathrm{min}$ all
  solutions are unbounded.
\item[iii.] For $|p_\eta > p_{\mathrm{crit}}|$ the local extrema of the
  effective potential have disappeared and it becomes a strictly
  decreasing function. In this case only unbounded solutions exist.
\end{itemize}
For negative chemical potential $\mu<0$ the effective potential is a
decreasing function and only unbounded solutions exist.

The bounded solutions for positive chemical potential can be extended
to global solutions on the infinite line straight forwardly. All
unbounded solutions develop a singularity at a finite value
$x_{\mathrm{sing}}$, where amplitude $R_{1,\pm 1}(x;r_0, p_{\eta},
H,\sigma )$ diverges like $1/|x-x_{\mathrm{sing}}|^{1/\nu}$ as can be
checked by inserting a wave function with that kind of singularity in
the stationary NLSE. Physically, such singularities indicate a
breakdown of the model as the NLSE is usually an effective description
of a physical system that is valid only for sufficiently small
amplitudes. Moreover, for $0<\nu\le 2$ the corresponding wave
functions are not square integrable over a finite interval containing
the singularity ($\int_{x_\mathrm{sing}}^{x_\mathrm{sing}+ \Delta
  x} \left| \Phi(x)\right|^2 dx \sim \int_0^{\Delta x} x^{-2/\nu} dx $
diverges).  For a Bose-Einstein condensate ($\nu=1$) this implies
infinitely many particles in a small interval around the singularity
which is also not physical. One is tempted to focus just on the
globally bounded solutions.  However, in the present setting we want
to use local solutions on finite intervals
to construct solutions on a graph; in that setting the globally unbounded
solutions cannot be excluded as they may still describe bounded
solutions on an edge of finite length (the singularity may only
develop on a larger distance).  It is not difficult to
construct global bounded solutions on a chain or ring graph which
involve any of the local solutions discussed above.

\subsubsection{The attractive case $g<0$}

\label{sec:attractive}
\begin{figure}[h]
  \begin{center}
    \includegraphics[width=0.45\textwidth]{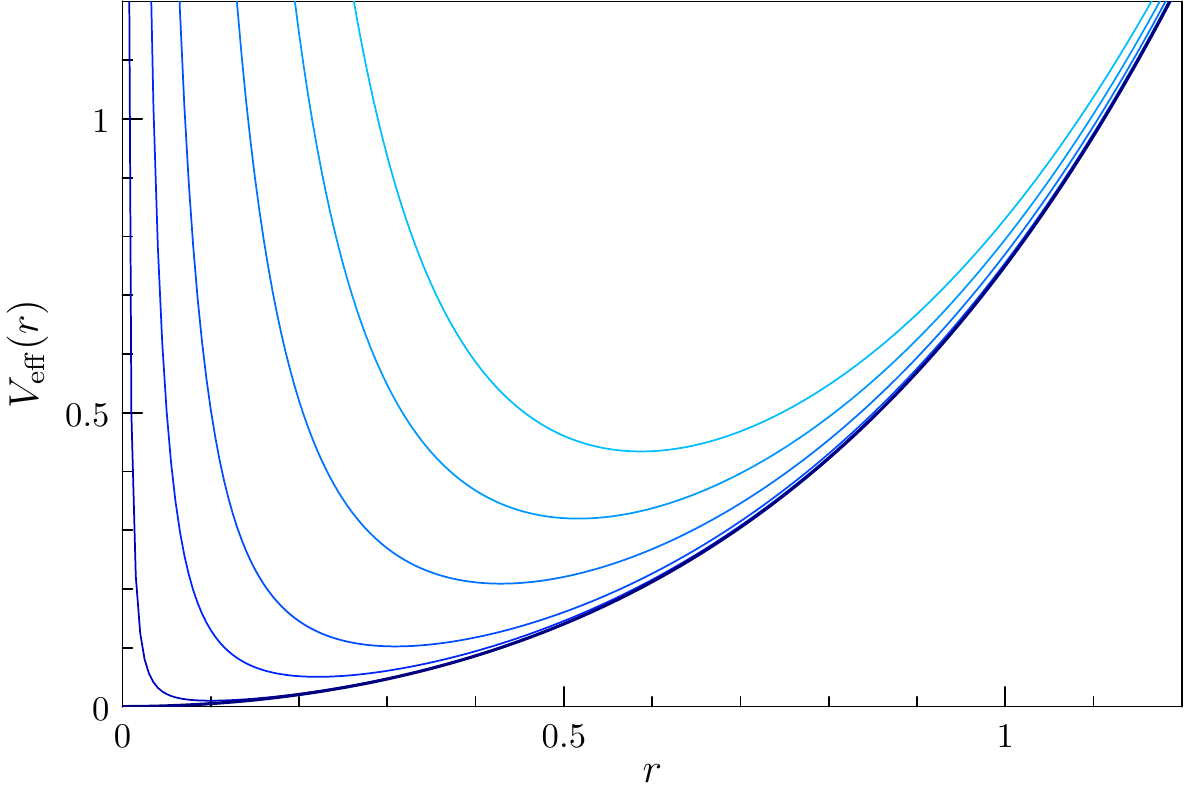}\hfill
    \includegraphics[width=0.45\textwidth]{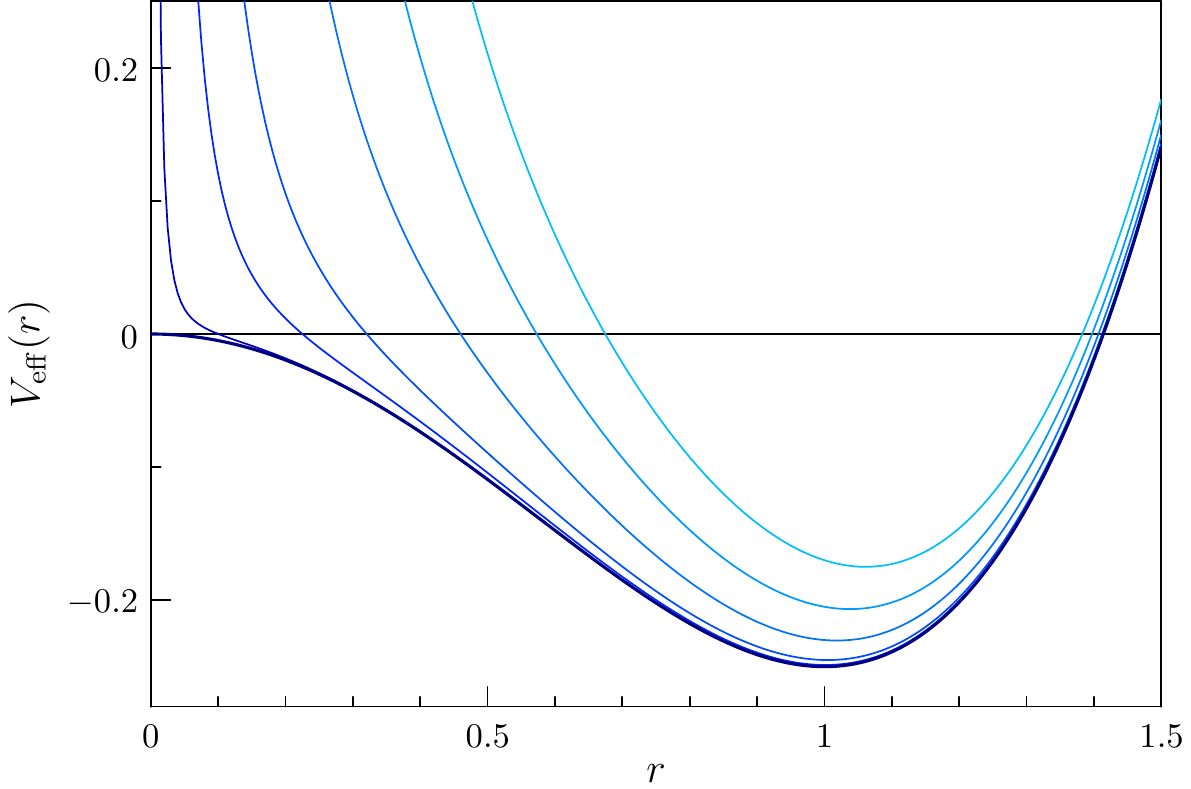}
    \caption{(Color Online) The effective potential $V_{\mathrm{eff}}(r)$  
      for the radial motion in the
      auxiliary dynamical system
      with attractive nonlinear coupling 
      $g=-1$ and $\nu=1$. In the left graph the
      chemical potential is $\mu=1$; 
      in the right graph it is $\mu=-1$.
      The curves in each graph correspond 
      to different values of the
      flow (angular momentum of auxiliary dynamics) 
      $p_\eta=(0,0.01,0.05,0.1,0.2,0.3,0.4)$.\\
      Graphs for different values of the 
      exponent $\nu$ look qualitatively similar.
      \label{fig:Veff_attractive}
    }
  \end{center}
\end{figure}

\begin{figure}[h]
  \begin{center}
    \includegraphics[width=0.45\textwidth]{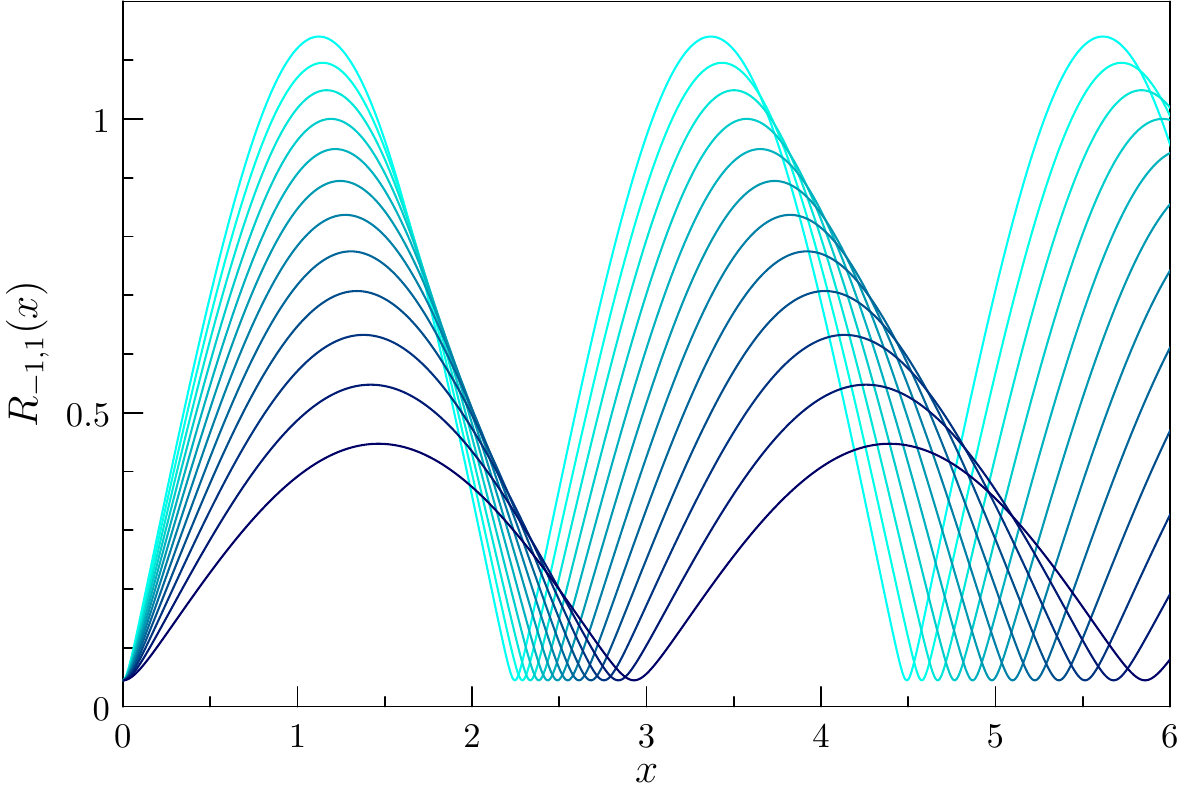}\hfill
    \includegraphics[width=0.45\textwidth]{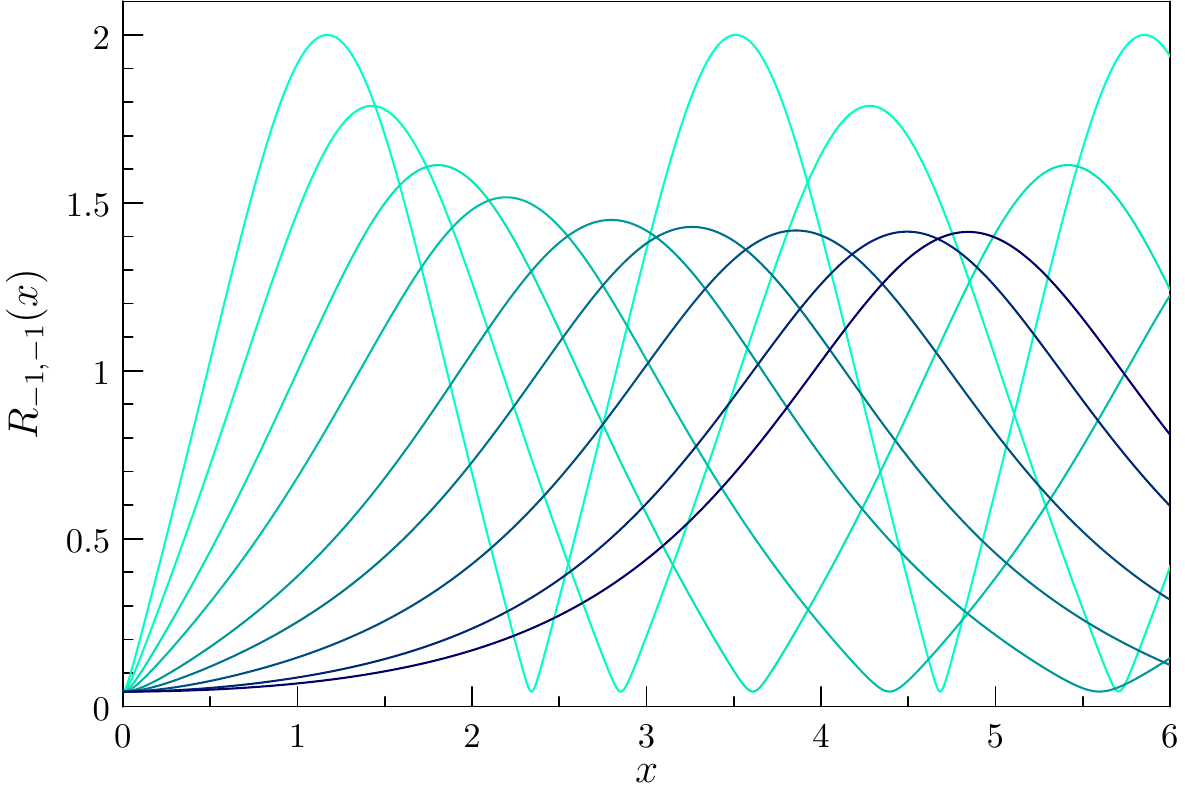}
    \caption{(Color Online) The amplitude functions $R_{-1,1}(x)$ (left) and
      $R_{-1,-1}(x)$ (right) for
      local solutions of the NLSE
      with repulsive interaction. 
      The parameters $H$, $p_\eta$ have been 
      varied and $r_0$ chosen such 
      that the minimum is at $x=0$.
      The plotted functions are
      for $\nu=1$ ($\nu \neq 1$ leads to qualitatively similar
      solutions).
      \label{fig:R_attractive}
    }
  \end{center}
\end{figure}

Here it is sufficient to consider
$g=-1$ and $\mu= \pm 1$. The effective
potential is
\begin{equation}
  V_{\mathrm{eff}}= 
  \frac{p_\eta^2}{2 r^2} \pm \frac{r^2}{2} 
  + \frac{1}{2\nu+2}r^{2\nu + 2}.
\end{equation}
Figure~\ref{fig:Veff_attractive} shows the effective potential for
various values of $p_\eta$.  Stationary solutions in the attractive
case always remain bounded with an amplitude $r_1 \le R_{-1,\pm 1}(x)
\le r_2$ (where the values of the turning points $r_1$ and $r_2$,
depend on $p_\eta$, $H$ and the sign of $\mu= \pm 1$).  The amplitude is a
periodic function $R_{-1,\pm 1}(x)=R_{-1,\pm 1}(x+\Lambda_{-1,\pm 1})$
with period
\begin{equation}
  \Lambda_{-1,\pm 1}(H,p_\eta)= 2 \int_{r_1}^{r_2}
  \left(2H -\frac{p_\eta^2}{r^2} 
    \mp r^2 - \frac{1}{\nu+1} r^{2\nu+2}\right)^{-1/2} dr\ .
\end{equation}
This period is finite unless $p_\eta=0$ and $H=0$ for a negative
chemical potential $\mu<0$. The solution in this case
consists of a single peak at $x=0$ which falls off exponentially on
both sides and then drops monotonically to zero as $|x| \to \infty$.
This solution is known as a stationary soliton \cite{Ablowitz}.
Figure \ref{fig:R_attractive} shows plots for the amplitude $R_{-1,\pm
  1}(x)$ for various
parameters including the soliton.\\
The phase function $\vartheta_{-1,\pm 1}(x)$
is again an increasing (decreasing) function if
$p_\eta>0$ ($p_\eta<0$) such that $\Delta \vartheta=
\vartheta_{-1,\pm 1}(x+\Lambda_{-1,\pm 1}) - \vartheta_{-1,\pm 1}(x)$
is in general not a rational multiple of $\pi$.  In the attractive
case the wave function $\phi(x)=R_{-1, \pm 1}(x) e^{i\vartheta_{-1,\pm
    1}(x)}$ is real if and only if there are nodal points 
on the infinite line (which is
equivalent to $p_\eta=0$).

\subsection{Stationary States for a Closed Graph as a Nonlinear
  Eigenproblem}

Let us now assume that we have a finite (connected) closed graph with
$E=B<\infty$ edges all of which are bonds. For the \emph{linear}
Schr\"odinger equation on such a graph (i.e. if the nonlinear coupling
constants vanish on each edge, $g_e=0$ for all $e \in \mathcal{E}$) it
is well known that stationary solutions only exist for a discrete set
of values $\mu_n$ of the chemical potential. In the context of the
linear Schr\"odinger operator these are the
energy eigenvalues of the system and the collection 
$\left\{\mu_n\right\}_{n=0}^\infty$ is the linear spectrum of the
graph. 
Spectral theory for linear quantum graphs is well developed
\cite{Kottos, GS_Adv, Berkolaiko}. Before developing the approach to nonlinear
quantum graphs, let us summarize how the linear spectrum is characterized as
the zero of an explicit characteristic function $\xi(k)$. 
Denoting a directed bond as a pair $(b,d)$, where $b$ denotes
the bond and $d=\pm 1$ is the direction such that $d=1$ ($d=-1$)
is the direction in which $x_b$ increases (decreases).
For positive chemical potential
$\mu=k^2>0$ the characteristic function has the form
\begin{equation}
  \xi(k)= \det(1-e^{ik \ell} S)
  \label{linear_quantization}
\end{equation} 
where $S$ is a unitary $2B \times 2B$ matrix that contains the quantum
amplitudes to scatter from one directed edge $(b,d)$ into another
directed edge $(b',d')$ and $e^{i k \ell}= \mathrm{diag} \left(
  e^{ik\ell_1}, \dots, e^{ik\ell_B}, e^{ik\ell_1}, \dots ,
  e^{ik\ell_B}\right)$ is a $2B \times 2B$ diagonal (unitary) matrix
that contains the phases $e^{ik \ell_b}$ that a plane wave acquires
going from one end of an edge to the other.  The matrix $S$ contains
information about the connectivity and about the matching
conditions.  
Its matrix elements $S_{bd, b'd'}$ vanish unless the
terminal vertex of the directed edge $(b',d')$ 
is the same as the
starting vertex of $(b,d)$. The nonvanishing values  $S_{bd, b'd'}$ depend
on the matching conditions.
Eq.~\eqref{linear_quantization} (or some relative of it) can be
used as the starting point for developing many tools of spectral
theory, such as trace formulas which express spectral functions
(e.g. the number of states inside a spectral interval) in terms of
periodic walks on the graph \cite{Kottos}.  
In the linear case one may always
normalize a solution such that $\left\|\Phi_n(x)\right\|^2=1$.\\
Let us now come back to 
the nonlinear case where stationary solutions can, in general, not be
multiplied by scalars and remain stationary solutions. Spectrum
and eigenfunctions of the linear quantum graph remain relevant for a
nonlinear quantum graph when the $L^2$-norm
\begin{equation}
  N= \left\|\Phi(x)\right\|^2
\end{equation}
is small, i.e., when we take $N \to 0$ we expect to recover the linear
spectrum and the linear eigenfunctions (up to normalization). For
arbitrary values of $N$ we see that stationary solutions exist
along one-parameter families of
values for the chemical potential $\mu_n(N)$.

This can be seen by constructing global solutions on the graph from
local solutions and adding matching conditions on the
graph.  If one first disregards any matching conditions, one may take the
values of the wave function $\phi_e(x_e)$ and its derivative at the
point $x_e=0$ as free parameters. The local solutions of the 
NLSE imply the local transfer operator that
allows us to find the corresponding values at the other end
of the edge $x_e=\ell_e$. 
This gives  $4B$ independent real parameters to which
we now add the matching conditions. Continuity at one vertex $i$ of
valency $v_i$ implies $2(v_i-1)$ real conditions; adding over all
vertices, this implies $4B-2V$ independent real conditions.  Next we
have the condition \eqref{matching_flux} for each vertex.  These give
together $2V-1$ independent real conditions. Note that one may have
expected $2V$ conditions; however, the imaginary part of the matching
condition \eqref{matching_flux} refers to flux conservation which is
also conserved by the solutions along the edge. So if the flux
condition is met at $V-1$ vertices it will automatically be conserved
at the last vertex as well. On the other side we are free to choose an
overall phase which adds one more condition. Note that
flux conservation is related to this global gauge symmetry by
Noether's theorem, so in a sense the `missing' condition we observed
for flux conservation at each vertex reappears in the form of one
parameter that we are free to choose. Altogether we have as many
conditions as free variables. So for given $\mu$ there will
generically be solutions for isolated points in the parameter space
that solve the problem. Each of these will have a definite value for
the norm $N$. If we fix $N$ from outside we have to leave $\mu$ as a
free parameter and will generically obtain solutions for discrete
values $\left\{ \mu_n(N)\right\}$ of the chemical potential.  As $N$
is changed the chemical potential of a given solution will change and
the corresponding wave function will deform. In principle, bifurcations
may occur when $N$ is changed; i.e., solutions may coalesce and
disappear, or solutions may appear.  In order to define a generalized
eigenvalue problem, one may fix $N$; in that case the nonlinear
spectrum $\left\{ \mu_n(N) \right\}_{n=0}^\infty$ will generally remain a
discrete set. Note that it is, in general, difficult to decide whether a
given set of generalized eigenvalues is the complete spectrum.  As
long as $N$ is sufficiently small, one may hope that the spectra
$\left\{ \mu_n(N) \right\}_{n=0}^\infty$ and the linear spectrum
$\left\{ \mu_n(0) \right\}_{n=0}^\infty$ are in one-to-one
correspondence and that eigenvalues with the same index $n$ are
continuously connected when
$N$ is changed from a finite value to zero.

Above we have used the values of the wave function and its derivatives
at $x_e=0$ on each bond, altogether $4B$ real parameters.  The
complexity of stating $4B$ equations for $4B$ parameters can be
reduced building in continuity at the vertices from the start, e.g by
choosing a spanning tree and then building up a continuous solution on
the spanning tree first. Further complexity reduction can be achieved
by considering the flux $p_\eta$ as a parameter on each edge;
flux conservation implies that it is sufficient to know the flux on
some edges (indeed just on the edges we took out to get a spanning
tree) to obtain the flux on other edges explicitly.

\subsection{Stationary Scattering States for an Open Graph}

Large parts of the discussion in the previous chapter can be extended
to open graphs with a finite number of leads. We just need to discuss
proper conditions for the wave function $\phi_\ell(x_\ell)$ on the
leads $\ell \in \mathcal{L}$ as $x_\ell \to \infty$. It is instructive
to first recall the situation for an open linear quantum graph where
$g_e=0$ for all edges $e\in \mathcal{E}$ \cite{Kottos2, Kottos3}. In
this case, basically two types of solutions exist: In the physics
literature they are usually referred to as bound states and scattering
states. The former have a discrete spectrum of eigenvalues and are
square-integrable, the latter have a continuous spectrum and have a
bounded amplitude on the leads (implying that they are not
square-integrable over the complete open graph).

\subsubsection{Bound states}

The bound state on a linear finite open graph have a discrete spectrum
of allowed values for the chemical potential.  The corresponding wave
function is either decaying exponentially $\phi_l \propto e^{-\kappa
  x_l}$ with a negative chemical potential $\mu=-\kappa^2<0$ (i.e., at
the bottom of the spectrum), or the wave function vanishes on all
leads. In the latter case there is a known topological mechanism that
makes it possible to construct wave functions with positive chemical potential
$\mu>0$ which are supported on finite closed subgraphs (topological
bound states). The topological bound states in a linear quantum graph
are not generic in the sense that they require rational ratios of bond
lengths.

In the nonlinear setting any bound states at the bottom of the
spectrum come in one-parameter families such that the chemical
potential depends on the number of particles ($L^2$-norm) in complete
analogy to the solutions in finite graphs.

The topological mechanism for bound states may also be generalized to
the nonlinear case; however, the number of conditions one has to impose
makes it clear that topological bound states remain as non-generic as
they are for the linear case (note that the condition of rational
ratios of bond lengths needs to be replaced with a non-linear
generalization).  A detailed understanding of topological bound states
for open nonlinear graphs is an interesting topic in its on right but
not in the focus of the present work which is either on closed graphs
or the scattering states of open graphs.

\subsubsection{Scattering states}

Let us start again with a discussion of the scattering states for a
linear open quantum graph.  These exist for any $\mu=k^2>0$ (the
continuous spectrum of a graph).  The wave function on the leads may
be written as
\begin{equation}
  \phi_l(x_l)= a_l e^{- i k x_l} + b_l e^{ikx_l},
  \label{lead_wf}
\end{equation}
where $a_l$ is the incoming amplitude and $b_l$ the outgoing amplitude
along the lead $l\in \mathcal{L}$. Physically, one may think of the
amplitudes $a_l$ as being fixed in an experiment and the outgoing
amplitudes $b_l$ as the response of the system that is to be
measured. Indeed, if one satisfies all matching conditions inside the
graph the outgoing amplitudes are related to the incoming by a linear
transformation
\begin{equation}
  b_l = \sum_{l'=1}^L S(k)_{l l'} a_{l'},
\end{equation}
where $S(k)$ is known as the \emph{scattering matrix}. Total flux
conservation implies that $S(k)$ is unitary. Explicitly the scattering
matrix of an open linear graph is given by
\begin{equation}
  S= \sigma_{\mathcal{LL}} +   \sigma_{\mathcal{LB}} \left(1-
    e^{ik \ell} \sigma_{\mathcal{BB}} \right)^{-1} 
  e^{i k \ell} \sigma_{\mathcal{LB}} 
\end{equation}
where $e^{ik\ell}=\mathrm{diag}(e^{ik \ell_1},\dots,e^{ik\ell_B},
e^{ik\ell_1},\dots, e^{ik \ell_B})$ and the matrices
$\sigma_{\mathcal{XY}}$ are constructed from the vertex scattering
matrices as follows. The diagonal $L \times L$ matrix
$\sigma_{\mathcal{LL}}$ contains all direct backscattering amplitudes
at the vertices adjacent to the corresponding lead. The $L \times 2B$
matrix $\sigma_{\mathcal{LB}}$ contains the scattering amplitudes for
scattering from a directed bond to an (outgoing) lead as the
corresponding entry in the matrix (matrix elements that are not
consistent with the directed edge connectivity vanish). Analogously,
the $2B \times L$ matrix $\sigma_{BL}$ contains the scattering
amplitudes for scattering from a lead (incoming) to a directed bond,
and the $2B \times 2B$ matrix $\sigma_{\mathcal{BB}}$ contains the
internal
scattering from one directed bond to another at some vertex.\\
The scattering matrix has a very clear physical interpretation in
terms of an experiment where incoming waves are fixed by the setting
and reflected or transmitted waves are measured. However, the incoming
wave and the reflected wave can only be characterized independently if
the
superposition principle holds on the leads.\\
For $\mu=k^2>0$ this can only be achieved by setting $g_l=0$ on all
leads $l\in \mathcal{L}$. With this assumption equation
\eqref{lead_wf} still describes the wave function on the leads such
that the coefficients $a_l$ are the amplitudes of the incoming wave
which we assume to be given by the experimenter (theoretically as
boundary conditions) and the coefficients $b_l$ describe the measured
response.  In general they are nonlinear functions,
\begin{equation}
  b_l= b_l(a_1,\dots,a_L) ,
\end{equation}
of the incoming amplitudes.  For sufficiently small incoming
amplitudes $|a_l|^2 \to 0$ one expects that the leading term is given
in terms of the scattering matrix of the corresponding linear graph
(i.e. setting $g_e=0$ everywhere on the graph)
\begin{equation}
  b_l(a_1,\dots,a_L) =\sum_{l'=1}^L S_{ll'}(k) a_{l'} + O(|a_1|^2,
  \dots, |a_L|^2)\ .
\end{equation}
Transmission and reflection can be defined if we have a single
non-vanishing incoming amplitude on the lead $l\in \mathcal{L}$. 
In that case we may define the reflection
coefficient as
\begin{equation}
  R_{l}(a_l)= \frac{|b_l(0,\dots,0,a_l,0,\dots)|^2}{|a_l|^2}
\end{equation}
and the transmission coefficient from lead $l$ to lead $l'$ as
\begin{equation}
  T_{l' l}(a_l)= \frac{|b_{l'}(0,\dots,0,a_l,0,\dots)|^2}{|a_l|^2}.
\end{equation}
Evaluating the functions $b_{l'}$ and the resulting reflection and
transmission coefficients is one of the central theoretical physical
problems for an open nonlinear quantum graph. Implicitly they are
given by the set of local solutions on the edges with the matching
conditions on the vertices as described before.\\
Let us conclude this section with two remarks.
\begin{itemize}
\item[i.] The assumption $g_l=0$ on the leads 
  is not a severe restriction for $\mu>0$. Indeed, if we have
  a scattering solution for a nonlinear open quantum graph where
  $g_l\neq 0$ for some lead $l \in \mathcal{L}$ then we may choose some
  point $x_{l,0}\ge 0$ on this lead and replace the solution for
  $x_l>x_{l,0}$ with \eqref{lead_wf} such that $\phi_l(x_l)$ and
  $\phi_l'(x_l)$ are continuous at $x_l=x_{l,0}$, while the solution for
  $x_l<x_{l,0}$ and all other edges remains unchanged. While the
  reflection and transmission amplitudes defined in this way will
  depend explicitly on the choice of $x_{l,0}$ (both the phase and the
  absolute values), such a construction allows us to discuss scattering
  solutions in analogy to linear wave scattering solutions.
\item[ii.]  While for linear open graphs the continuous
  spectrum is always $\mu\ge 0$ it is possible to have scattering solutions
  on nonlinear open graphs where $\mu<0$ (if $g_l<0$ on some leads). In
  that case we may not just set $g_l=0$ on all leads without changing
  the scattering solution on the bonds unless we are ready to accept
  solutions that grow in absolute value without bound along the leads.
  However, if we introduce a constant negative potential $-V_0<0$ along
  the leads such that $\mu+V_0=k^2>0$, we may set $g_l=0$ in an analogous
  way and then discuss scattering solutions in analogy to the case
  $\mu>0$.
\end{itemize}

\section{Hamiltonian Formalism and Canonical 
  Perturbation Theory for
  the NLSE}
\label{canon_pert}

The framework presented in the previous section makes it possible to reduce the
problem of finding solutions to the NLSE on graphs to a finite set of
(generally nonlinear) algebraic equations for a finite set of variables.
Already for relatively simple graphs the complexity of solving these
equations will not allow for explicit analytical solutions. Compared
to the corresponding problem of finding solutions to the linear
Schr\"odinger equations on graphs also a numerical approach is faced
with a considerably increased complexity. In the second paper \cite{GW2} 
of this series we will apply the framework to find 
some solutions for a few 
basic closed and open graph structures. In the face of the rising
complexity of the problem when the number of edges grows we
introduce an approximation scheme that assumes small
amplitudes and allows for long edges $\ell_b \gg 1/k$ with a positive chemical
potential $\mu=k^2>0$.  
This approximation scheme is based on standard
canonical perturbation theory for the auxiliary Hamiltonian dynamics
as described in textbooks \cite{Lichtenberg, Tabor}. We start with
formally defining exact action-angle variables for the system. These
depend formally on the nonlinear coupling strength $g$. For $g\to 0$
the action-angle variables reduce to the well-known action angle
variables of a two-dimensional isotropic harmonic oscillator. The
exact action-angle variables at finite $g$ can then be expressed as
a formal expansion in $g$ using canonical perturbation theory.
We derive explicitly to lowest order in the cubic case $\nu=1$.

There are several advantages of the canonical perturbation theory
over a direct expansion of the wave function in the NLSE. In such an
approach one sets
$\phi(x)=\phi^{(0)}(x)+\delta\phi(x)$ where $\phi^{(0)}(x)$ is a
solution of the linear Schr\"odinger equation and
$\delta\phi(x)=\sum_{n=1}^\infty g^n \phi^{(n)}(x)$ 
accounts for the perturbation.
Then one solves the
equations order by order. In the most naive variant it is well known
from the standard textbook example \cite{Tabor} that
unphysical resonance effects increase the amplitude of oscillations
in  $\delta \phi(x)$ effectively destroying the applicability of the
whole approach at a finite distance. More sophisticated variants of
this approach (e.g. by adding formal expansion of other parameters,
e.g. setting $\mu=\mu^{(0)}+\sum_{n=1}^\infty g^n \mu^{(n)}$) may
improve this but will to lowest order always keep the form
$\phi(x)\approx\phi^{(0)}(x)+ g\phi^{(1)}(x)$ where $g\phi^{(1)}(x)$
is supposed to be a small perturbation of the leading term for
arbitrary (large) $x$.\\ 
The main root of the unphysical (and mathematically unwanted) 
resonance effects lies in the fact that the wavenumber changes when
the system is perturbed. The unperturbed wave function is periodic
and obeys $\phi(x)=\phi(x+2\pi/k)$. The perturbed solutions are 
quasi-periodic with two (generally
incommensurate for finite $g$) wave numbers. Using action-angle variables
is the natural way to decouple these two underlying periods.
Moreover this approach shows that the perturbations lead to two
different
effects: They change the local shape of the
wave function and they change the wave numbers. 
While tiny changes of the shape are locally confined, even the most
tiny shifts in the wavenumber lead to changes of the corresponding
phases
that can add up over large distances such that the wave function can
no longer be written in the form $\phi(x)\approx\phi^{(0)}(x)+
g\phi^{(1)}(x)$
in any consistent way.
Indeed while we derive canonical perturbation theory formally as an expansion
in the nonlinear coupling strength $g$ the approach opens a number
of asymptotic regimes in the cubic NLSE that we discuss at the end of the section.

\subsection{Hamiltonian Formalism and Action-Angle Variables}

As we have described the auxiliary dynamics in the Lagrangian approach
it is straight forward to perform the standard Legendre transform and
the corresponding change of variables $(r,\eta,dr/dx,d\eta/dx)
\mapsto (r,\eta,p_r,p_\eta)$ where $p_r =dr/dx$ is the conjugate
momentum to the variable $r$ and $p_\eta$ defined in
\eqref{eq:angular_momentum} is the angular momentum conjugate to
$\eta$.  
The Hamilton function is just the energy \eqref{eq:energy}
expressed in the canonical variables
$  H= \frac{1}{2} p_r^2 + V_{\mathrm{eff}}(p_\eta , r)$.
As this is an integrable system the Hamiltonian equations of motion are
simplified by introducing action-angle variables.  We assume a positive
chemical potential $\mu >0$ for the rest of this section. This ensures
oscillatory solutions for sufficiently small  $r$ and $p_r$. 
This is the region where we want to define action-angle
variables.  In the attractive case $g<0$ the approach is valid in the
whole phase space. \\
In the present context one action variable is the angular momentum
\begin{equation}
  I_\eta\equiv p_\eta
\end{equation}
which can take any value in $\mathbb{R}$. 
The second action variable can be expressed as the integral
\begin{equation}
  I_r(H,I_\eta)=\frac{1}{\pi}\int_{r_-(H,I_\eta)}^{r_+(H,I_\eta)}
  \sqrt{2(H-V_{\mathrm{eff}}(I_\eta,r))} dr
  \label{eq:radial_action}
\end{equation}
which is expressed as a function of the angular momentum and
energy.
Here
$r_+(H,I_\eta)$ and $r_-(H,I_\eta)<r_+(H,I_\eta)$ are the
turning points defined as solutions of $V_{eff}(I_\eta,
r_\pm)=H$. Note that \eqref{eq:radial_action} implies $I_r\ge 0$ and
implicitly defines the energy as a function of the action variables
$H=H(I_r,I_\eta)$.  Moreover, by expressing $H=p_r^2/2+
V_{\mathrm{eff}}(p_\eta,r)$ Eq.~\eqref{eq:radial_action} also defines
the radial action as a function of the original phase space
coordinates, i.e.
$I_r=I_r(p_r,p_\eta,r)$.\\
This allows us to define a generating function for a canonical
transformation $(r,p_r,\eta,p_\eta)\mapsto
(\alpha_r,I_r,\alpha_\eta,I_\eta)$.  We can write the generating
function that depends on the original (generalised position) variables
$r$ and $\eta$ and the new action variable $I_r$ and $I_\eta$
\begin{equation}
  S(I_r,r,I_\eta,\eta)=I_\eta \eta + \int^r_{r_0} 
  p_r(I_r, I_\eta,r') dr' + F(I_r,I_\eta),
  \label{eq:generating_function}
\end{equation}
where $p_r(I_r, I_\eta,r)$ is defined implicitly by
\eqref{eq:radial_action} and expressing $H=H(p_r,I_\eta,r)$. The
function $F(I_r, I_\eta)$ can be chosen arbitrarily as it only
affects a shift of the angle variables.  The lower boundary $r_0$ of
the integral in \eqref{eq:generating_function} is an arbitrary
constant (in principle one may incorporate the effect of
$F(I_r,I_\eta)$ into the lower boundary by letting it depend on the
actions).  The transformation is generated by taking derivatives of
\eqref{eq:generating_function}
\begin{subequations}
  \begin{align}
    p_r=&\frac{\partial S}{\partial r} = p_r(I_r,I_\eta,r)\\
    p_\eta=& \frac{\partial S}{\partial \eta} =I_\eta\\
    \alpha_r=& \frac{\partial S}{\partial I_r} = \frac{\partial
      F(I_r,I_\eta)}{\partial I_r}+ \int^r_{r_0}
    \frac{\partial p_r(I_r,I_\eta,r')}{\partial I_r} dr'\\
    \alpha_\eta=&\frac{\partial S}{\partial I_\eta} = \eta +
    \frac{\partial F(I_r,I_\eta)}{\partial I_\eta}+ \int^r_{r_0}
    \frac{\partial p_r(I_r,I_\eta,r')}{\partial I_\eta} dr'\ .
  \end{align}
\end{subequations}
Here the first equation gives back \eqref{eq:radial_action} and the
second gives $p_\eta=I_\eta$, as required. The third and fourth
equations define
the two angle variables $\alpha_r$ and $\alpha_\eta$.\\
The complexity of the transformation to action-angle variables is
accompanied by a corresponding simplification of the equations of
motion. By construction the two action variables are constants of
motion and the angle variables change linearly in time
\begin{equation}
  \alpha_r(t)=\alpha_r(0) + \kappa_r(I_r,I_\eta)t
  \qquad \text{and} \qquad
  \alpha_\eta(t)=\alpha_\eta(0) + \kappa_\eta(I_r,I_\eta)t
\end{equation}
where
\begin{equation}
  \kappa_r= \frac{\partial H}{\partial I_r}
  \qquad
  \text{and} \qquad
  \kappa_\eta= \frac{\partial H}{\partial I_\eta}\ .
\end{equation}
These two angular frequencies (or wave numbers in the original context) may be given in a slightly more explicit
way as
\begin{equation}
  \kappa_r=\frac{1}{\frac{\partial I_r}{\partial H}}
  \qquad \text{and} \qquad
  \kappa_\eta= - 
  \frac{\frac{\partial I_r}{\partial I_\eta}}{\frac{\partial I_r}{\partial H}}
  \label{eq:wavenumbers_exact}
\end{equation}
in terms of the function $I_r(H,I_\eta)$ as
defined in \eqref{eq:radial_action}. For the cubic case $\nu=1$ we  give exact expressions for 
$\partial I_r/\partial H$ and 
$\partial I_r/\partial I_\eta$ in Appendix~\ref{angularfrequencies}.\\
While the Hamiltonian description in action-angle variables does not
seem to simplify the full solution of the problem it is a
reformulation that will allow us to perform a systematic
expansion that, at least in low orders, offers closed analytic expressions.\\
Our strategy will be to find approximate solutions to the
transformation $(r,p_r,\eta,I_\eta)\mapsto
(\alpha_r,I_r,\alpha_\eta,I_\eta)$ by \textit{formally} considering the nonlinear
coupling constant $g$ as a small parameter.  For this we write the
original Hamiltonian as
\begin{equation}
  H(p_r,I_\eta,r)=H_0(p_r,I_\eta,r) - g \frac{r^{2\nu+2}}{2\nu+2}.
\end{equation}
where $H_0(p_r,I_\eta,r)$ is the Hamilton function of the linear
problem.  We start with transforming to action-angle variables of the
linear case and then use canonical perturbation theory to treat the
additional term $g \frac{r^{2\nu+2}}{2\nu+2}$. Such a perturbative
treatment is valid as long as the harmonic term in the effective
potential dominates the anharmonic perturbation, that is $k^2
\frac{r^2}{2} \gg g \frac{r^{2\nu+2}}{2\nu+2} $ or
\begin{equation}
  g \frac{r^{2\nu}}{(\nu+1) k^2} \ll 1\ .
  \label{eq:perturbation_parameter}
\end{equation}
As $\nu>0$ the perturbative expansion will only be valid for small
amplitudes and break down as soon as amplitudes are of size $r^2 \sim
(k^2/g)^{1/\nu}$.

\subsection{The linear case}

If $g=0$ the transformation to action-angle coordinates can be
performed explicitly. In order to distinguish the action-angle
variable for $g=0$ from the exact action-angle variable for $g\neq 0$,
we use the variables $(\beta_r,J_r,\beta_\eta,J_\eta)$ for
$g=0$ and reserve $(\alpha_r,I_r,\alpha_\eta,I_\eta)$ for the
exact action-angle variables in the general case. With
$J_\eta=p_\eta$ one may perform the corresponding integral in
\eqref{eq:radial_action} to obtain
\begin{equation}
  J_r(H_0,J_\eta)=\frac{H_0}{2k}-\frac{\left|J_\eta\right|}{2}
  \quad \Leftrightarrow \quad
  H_0(J_r,J_\eta)=k\left(2 J_r + \left|J_\eta\right|  \right)\ .
\end{equation}
In the following we always consider $H_0$ as a function of $J_r$ and
$J_\eta$.  The generating function \eqref{eq:generating_function}
for the transformation can be expressed explicitly as
\begin{equation}
  S_0(J_r,J_\eta,r,\eta)=J_\eta \eta +\frac{H_0}{2k}
  \left[a\sqrt{1-u^2} +\arcsin(u)- \sqrt{1-a^2}\arcsin\left(\frac{a+u}{1+au}\right)\right]
\end{equation}
where
\begin{subequations}
  \begin{align}
    a=& \sqrt{1-\frac{J_\eta^2 k^2}{H_0^2}} =
    \frac{2\sqrt{J_r (J_r+|J_\eta|)}}{2J_r+|J_\eta|}\\
    u=&\frac{\frac{k^2 r^2}{H_0(J_r,J_\eta)}-1}{a} \ .
  \end{align}
\end{subequations}
The resulting angle variables are
\begin{subequations}
  \begin{align}
    \beta_r=& \arcsin(u)\\
    \beta_\eta=& \eta- \frac{s_\eta}{2} \arccos\left(1-
      \frac{2J_r(1-u^2)}{(2J_r+|J_\eta|)(1+au)} \right)
  \end{align}
\end{subequations}
where $s_\eta= \mathrm{sgn}(J_\eta)=J_\eta/|J_\eta|$.\\
Hamilton's equations in the action-angle variables leave the action
variables $J_r$ and $J_\eta$ constant, while the angles change
linearly
\begin{equation}
  \beta_r(x)= 2k x +\beta_r(0)
  \quad \text{and} \quad
  \beta_\eta(x)= s_\eta k x + \beta_\eta(0)\ .
  \label{eq:angle_solution_linear}
\end{equation}
The wave function $\phi(x)=r(x)e^{i \eta(x)}$ can be expressed via
\begin{subequations}
  \begin{align}
    r(x)=&
    \sqrt{\frac{2J_r+|J_\eta|}{k}\left(1+a\sin(\beta_r(x))\right)}
    \label{eq:amplitude_actionangle}
    \\
    \eta(x)=&\beta_\eta(x)+ \frac{s_\eta}{2} \arccos\left(1-
      \frac{2J_r\cos^2(\beta_r(x))}{(2J_r+|J_\eta|)(1+a\sin(\beta_r(x)))}
    \right)
    \label{eq:phase_actionangle}
    \\
    \phi(x)= & \frac{1}{\sqrt{k}} \left( \sqrt{J_r+|J_\eta|} e^{i
        \beta_\eta(x)} + i s_\eta \sqrt{J_r} e^{i
        \beta_\eta(x)-i s_\eta \beta_r(x)} \right)
    \label{eq:wavefunction_actionangle}
  \end{align}
\end{subequations}
in terms of action-angle variables. Using $s_\eta=\pm 1$ and the
solutions \eqref{eq:angle_solution_linear} (with vanishing initial
angles) the exponentials in \eqref{eq:wavefunction_actionangle} reduce
to the well known solutions $e^{i\beta_\eta(x)}= e^{\pm i k x}$ and
$e^{i\beta_\eta(x) -i s_\eta \beta_r(x)}=e^{\mp i s_\eta k
  x}$.  Note also that \eqref{eq:amplitude_actionangle} implies
\begin{equation}
  \frac{\left(\sqrt{J_r+|J_\eta|}-\sqrt{J_r}\right)^2}{k} \le r(x)^2
  \le
  \frac{\left(\sqrt{J_r+|J_\eta|}+\sqrt{J_r}\right)^2}{k}
  \label{eq:intensity_bounds}
\end{equation}
and $r(x)^2=|\phi(x)|^2$ oscillates betweens these bounds with a
wavelength $\pi/k$.
\\
While \eqref{eq:wavefunction_actionangle} seems a complicated way to
write a quite trivial solution it is the starting point of the
canonical perturbation theory that will take into account the
nonlinearity.  In the regime of weak nonlinearity, i.e.  $g r^{2\nu}
\ll k^2$ we will see that the accumulated effect of the nonlinearity
over large distances can be captured to leading order by keeping the
form of the wave function \eqref{eq:wavefunction_actionangle} and
replacing $k$ in \eqref{eq:angle_solution_linear} with two perturbed
wave numbers $\kappa_r(J_r,J_\eta)$ and
$\kappa_\eta(J_r,J_\eta)$ which will depend on the action
variables.

\subsection{Canonical perturbation theory}

After having found action-angle variables $(\beta_r, J_r,
\beta_\eta, J_\eta)$ for the linear case let us now write the
full nonlinear Hamilton function in terms of these action-angle
variable
\begin{equation}
  H(J_r,J_\eta,\beta_r)=H_0(J_r,J_\eta) + g G_0(J_r,J_\eta,\beta_r)
\end{equation}
where
\begin{equation}
  G_0(J_r,J_\eta,\beta_r)=
  -\frac{\left(2J_r+|J_\eta|
      + 2\sqrt{J_r(J_r+|J_\eta|)}
      \sin(\beta_r)\right)^{\nu+1}}{k^{\nu+1}(2\nu +2)}
\end{equation}
The perturbative parameter which is considered small in the following
is $g r^{2\nu}/k^2 \ll 1 $ (see \eqref{eq:perturbation_parameter}).
Using \eqref{eq:intensity_bounds} this is equivalent to requiring
\begin{equation}
  \frac{g (J_r+|J_\eta|)^\nu}{k^{2+\nu}} \ll 1
\end{equation}
or $g\ll k^{2+\nu}/(J_r+|J_\eta|)^\nu$.\\
In $n$-th order perturbation theory we want to find transformed
action-angle variables $(\alpha_r^{(n)}, I_r^{(n)},
\alpha_\eta^{(n)}, I_\eta^{(n)})$ such that the Hamilton
function expressed in new variables becomes
\begin{equation}
  H=H_0(J_r,J_\eta) + g G_0(J_r,J_\eta,\beta_r) 
  \equiv H_n(I_r^{(n)},I_\eta^{(n)}) +
  g^{n+1} G_{n+1}(I_r^{(n)},I_\eta^{(n)},\beta_r)
  \label{eq:reduced_Hamiltonfunction}
\end{equation} 
where $\beta_r=\beta_r(I_r^{(n)},I_\eta^{(n)},\alpha_r^{(n)})$.
For $n=0$ we set $I_r^{(0)}=J_r$, $I_\eta^{(0)}=J_\eta$,
$\alpha_r^{(0)}=\beta_r$ and $\alpha_\eta^{(0)}=\beta_\eta$. To
find the generating function of the canonical transformation, one uses
the ansatz
\begin{equation}
  S_n(I_r^{(n)},I_\eta^{(n)},\beta_r,\beta_\eta)=
  I_r^{(n)}\beta_r+  I_\eta^{(n)}\beta_\eta
  +\sum_{m=1}^n  g^m F_m(I_r^{(n)},I_\eta^{(n)},\beta_r)
\end{equation} 
where the functions $F_m(I_r^{(n)},I_\eta^{(n)},\beta_r)$ are
periodic in $\beta_r$ and found by the requirement that the generated
transformation
\begin{subequations}
  \begin{align}
    J_r=& \frac{\partial S_n}{\partial \beta_r}= I_r^{(n)}+
    \sum_{m=1}^n g^m \frac{\partial F_m}{\partial \beta_r}\\
    J_\eta=&\frac{\partial S}{\partial \beta_\eta}=
    I_\eta^{(n)}\\
    \alpha_r^{(n)}=&\frac{\partial S}{\partial I_r^{(n)}} =\beta_r
    +\sum_{m=1}^n g^m \frac{\partial F_m}{\partial
      I_r^{(n)}}\\
    \alpha_\eta^{(n)}=&\frac{\partial S}{\partial I_\eta^{(n)}}
    =\beta_\eta +\sum_{m=1}^n g^m \frac{\partial F_m}{\partial
      I_\eta^{(n)}}
  \end{align}
\end{subequations}
leads to the cancellation of all terms involving $\beta_r$ up to
$n$-th order in \eqref{eq:reduced_Hamiltonfunction}. This can be done
in an iterative manner by expanding $H_0(I_r^{(n)}+ \sum_{m=1}^n g^m
\frac{\partial F_m}{\partial \beta_r}, I_\eta^{(n)}) $ and
$G_0(I_r^{(n)}+ \sum_{m=1}^n g^m \frac{\partial F_m}{\partial
  \beta_r}, I_\eta^{(n)},\beta_r) $ in orders of $g$ and demanding
that the terms cancel order by order. If all $F_m$ for $m\le n$ are
found, one can immediately proceed to the order $n+1$ in the
perturbation theory where the known functions $F_m$ may be kept and
only $F_{n+1}$ needs
to be found.\\
In first order perturbation theory one finds
\begin{equation}
  H_0+g G_0= k\left(2I_r^{(1)} +|I_\eta|^{(1)} \right) + 2g
  k\frac{\partial F_1}{\partial \beta_r} -\frac{g}{k^{\nu+1}(2\nu +2)} 
  \left(2I_r^{(1)}+|I_\eta^{(1)}|
    +2\sqrt{I_r^{(1)}(I_r^{(1)}+|I_\eta^{(1)}|)}\sin(\beta_r)
  \right)^{\nu+1} + O(g^2)\ .
  \label{eq:firstoder_general}
\end{equation}
This can be solved in principle for any $\nu>0$ by writing
$F_1(I_r^{(1)},I_\eta^{(1)},\beta_r)=\sum_{N=-\infty}^\infty
f_{1N}(I_r^{(1)},I_\eta^{(1)}) e^{i\beta_r N}$ and requiring that
$i k N f_{1N}(I_r^{(1)},I_\eta^{(1)})$ cancels the corresponding
Fourier coefficient of $g G_0$.\\
Let us here focus on the most relevant case of the cubic nonlinearity
($\nu=1$), where \eqref{eq:firstoder_general} reduces to
\begin{equation}
  \begin{split}
    H_0+g G_0= &k\left(2I_r^{(1)} +|I_\eta|^{(1)} \right) -
    \frac{g}{4k^2}\left({6I_r^{(1)}}^2+ 6 I_r^{(1)}|I_\eta^{(1)}| +
      {I_\eta^{(1)}}^2\right)
    \\
    &+2g k\frac{\partial F_1}{\partial \beta_r} - \frac{g}{2k^{2}}
    \left(
      2(2I_r^{(1)}+|I_\eta^{(1)}|)
      \sqrt{I_r^{(1)}(I_r^{(1)}+|I_\eta^{(1)}|)}\sin(\beta_r)
      -I_r^{(1)}(I_r^{(1)}+|I_\eta^{(1)}|)\cos(2\beta_r)
    \right)+O(g^2)
  \end{split}
  \label{eq:firstoder_cubic}
\end{equation}
where the term in the second line cancels by choosing
\begin{equation}
  F_1=
  -\frac{1}{2 k^3} (2I_r^{(1)}+|I_\eta^{(1)}|)
  \sqrt{I_r^{(1)}(I_r^{(1)}+|I_\eta^{(1)}|)}\cos(\beta_r)
  -
  \frac{1}{8 k^3}I_r^{(1)}(I_r^{(1)}+|I_\eta^{(1)}|)\sin(2\beta_r)\ .
\end{equation}
and the Hamilton function is $H=H_1(I_r^{(1)},I_\eta^{(1)})+ g^2
G_1(I_r^{(1)},I_\eta^{(1)},\beta_r)$ with
\begin{equation}
  H_1(I_r^{(1)},I_\eta^{(1)})=
  k\left(2I_r^{(1)} +|I_\eta^{(1)}| \right) -
  \frac{g}{4k^2}
  \left({6I_r^{(1)}}^2+ 6 I_r^{(1)}|I_\eta^{(1)}| + {I_\eta^{(1)}}^2\right)\ .
  \label{eq:Hamilton_firstorder}
\end{equation}
Neglecting quadratic orders in $g$, the original action-angle variables
can be expressed in terms of the new ones as
\begin{subequations}
  \begin{align}
    J_r=& I_r^{(1)} +\frac{g}{4k^3}\left(
      2(2I_r^{(1)}+|I_\eta^{(1)}|)
      \sqrt{I_r^{(1)}(I_r^{(1)}+|I_\eta^{(1)}|)}\sin(\alpha_r^{(1)})
      - I_r^{(1)}(I_r^{(1)}+|I_\eta^{(1)}|)\cos(2\alpha_r^{(1)})
    \right)
    \\
    J_\eta=& I_\eta^{(1)}\\
    \beta_r=& \alpha_r^{(1)} +\frac{g}{8k^3} \left(
      \frac{16{I_r^{(1)}}^2+
        16I_r^{(1)}|I_\eta^{(1)}|+2{I_\eta^{(1)}}^2}{
        \sqrt{I_r^{(1)}(I_r^{(1)}+|I_\eta^{(1)}|)}}
      \cos(\alpha_r^{(1)}) +(2I_r^{(1)}+|I_\eta^{(1)}|)\sin(2
      \alpha_r^{(1)})
    \right)\\
    \beta_\eta=&\alpha_\eta^{(1)} +s_\eta\frac{g}{8 k^3}
    \left( \frac{8{I_r^{(1)}}^2+ 6I_r^{(1)}|I_\eta^{(1)}|}{
        \sqrt{I_r^{(1)}(I_r^{(1)}+|I_\eta^{(1)}|)}}
      \cos(\alpha_r^{(1)}) +I_r^{(1)}\sin(2 \alpha_r^{(1)}) \right)
  \end{align}
  \label{eq:first_order_transformation}
\end{subequations}
where $s_\eta= \mathrm{sgn}(I_\eta^{(1)})$. \\
The solution of the Hamiltonian dynamics in first order perturbation
theory leaves the action variables $I_r^{(1)}$ and $I_\eta^{(1)}$
constant, while the conjugate angles increase as
\begin{subequations}
  \begin{align}
    \alpha_r^{(1)}(x)=& \kappa_r^{(1)}(I_r^{(1)},I_\eta^{(1)}) x
    +\alpha_r^{(1)}(0) \\
    \alpha_\eta^{(1)}(x)=&
    \kappa_\eta^{(1)}(I_r^{(1)},I_\eta^{(1)}) x
    +\alpha_\eta^{(1)}(0)
  \end{align}
  \label{eq:angle_solution}
\end{subequations}
where
\begin{subequations}
  \begin{align}
    \kappa_r^{(1)}(I_r^{(1)},I_\eta^{(1)}) =& \frac{\partial
      H_1}{\partial I_r^{(1)}}= 2k\left(1 -\frac{3g}{4k^3}(2
      I_r^{(1)}+ |I_\eta^{(1)}|)\right)
    \\
    \kappa_\eta^{(1)}(I_r^{(1)},I_\eta^{(1)}) =&\frac{\partial
      H_1}{\partial I_\eta^{(1)}}= s_\eta k\left(1 -
      \frac{g}{2k^3} (3 I_r^{(1)} + |I_\eta^{(1)}|) \right) \ .
  \end{align}
\end{subequations}
Substitution of this solution into \eqref{eq:wavefunction_actionangle}
gives an approximate local solution for the stationary 
NLSE to first order in
$\frac{g(I_r^{(1)}+|I_\eta^{(1)}|)}{k^3}\ll 1$.  

\subsection{The asymptotic regimes of the nonlinear transfer operator
in canonical perturbation theory}

In order to understand the significance of the approximate solutions to 
the NLSE in canonical perturbation theory to lowest nontrivial order,
let us consider the nonlinear transfer operator along some edge of length 
$\ell$ in the graph. For this purpose
it is not necessary to consider the full nonlinear transfer operator 
in its most general form. It will be 
sufficient to restrict the `initial' conditions at $x=0$
to $\phi(0)=0$ (while the derivative takes some real value)
and only consider the wave function at the other end of the edge.
In first-order canonical perturbation theory, this is given  by
\begin{subequations}
  \begin{align}
    \phi(\ell)=&2 
                 \sqrt{\frac{J_r(\ell)}{k}}
                 \sin\left(
                 \frac{\beta_r(x)}{2}
                 \right),\\
    J_r(\ell)=&I_r-\frac{gI_r^2}{4k^3}
                \left(4
                \cos(\alpha_r(\ell))-\cos(2\alpha_r(\ell))\right)+O\left(\frac{g^2I_r^3}{k^6}\right),\\
    \beta_r(\ell)=&\alpha_r(\ell)
                    +\frac{gI_r}{4k^3}\left( 8\sin(\alpha_r(\ell))-\sin(2\alpha_r(\ell))\right)
                    +O\left(\frac{g^2 I_r^2}{k^6}   \right),\\
    \alpha_r(\ell)=& \kappa_r \ell,\\
    \kappa_r=& 2k\left(
               1-\frac{3gI_r}{2k^3}+
               O\left(
               \frac{g^2 I^2}{k^6}
               \right)
               \right)\ .
  \end{align}
  \label{eq:real_perturbation_sol}
\end{subequations}
This is a real solution where only one action-angle pair is relevant.
While this special case does not show the dephasing between the two degrees 
of freedom that are present for more general initial conditions, the
explicitly given error estimates will be sufficient to identify 
the relevant asymptotic regimes and
these regimes remain unaltered in the general case.
Equations
\eqref{eq:real_perturbation_sol} are justified  
for locally weak nonlinearity which really means that the dimensionless
strength of nonlinearity is negligible
$\frac{|g| \overline{|\phi|^2}}{k^2}\propto \frac{gI_r}{k^3}\ll 1$. 
Equations
\eqref{eq:real_perturbation_sol} also 
reveal two entirely different effects of a weak nonlinearity on a
solution. 
The first effect  is a local deformation 
of the linear
solution. The second effect is a
phase shift due to the nonlinear
wave numbers $\kappa_r$.  
Our approach assumed locally weak nonlinearity which
implies that the local deformations (which are of relative order 
$\frac{|g|I_r}{k^3}$) 
are always small. However 
the accumulated change in the phase 
(which is of order $
\frac{|g|I_r \ell}{k^2}$) 
does not necessarily need to be small.\\ 
One may identify three different asymptotic regimes
that are consistent with the canonical perturbation
expansion. 
Each may lead to additional consistent simplifications.
\begin{itemize}
\item[R1] The \emph{low-intensity weakly nonlinear asymptotic regime} 
  $g \overline{\left|\phi\right|^2} \to 0$ at fixed (bounded) 
  wavenumber $k$. This is equivalent to either $g \to 0$ or
  $I_r \to 0$ (and $I_\eta\to 0$ for general initial conditions) 
  when all other parameters are fixed. 
  This regime is weak in both the local and the global sense.
  For the leading nonlinear effects one may expand
  the oscillatory functions with respect to the 
  small phase shifts (where this leads to a simplification). 
\item[R2] The \emph{short wavelength globally weak nonlinear asymptotic regime}
  $k \to \infty$ with $g \overline{\left|\phi_e\right|^2}$ fixed (bounded).
  This is similar to the low-intensity regime in that it is 
  weakly nonlinear in both the local and the global sense. 
  It leads to additional
  simplifications as the dominant nonlinear effects all come from
  the shift in the nonlinear wave number $\kappa_r$ 
  (and $\kappa_\eta$ for general initial conditions).
\item[R3]
  The \emph{short wavelength asymptotic regime with moderately 
    large intensities}
  $ k \to \infty$ and $\frac{g \overline{|\phi|^2}}{k^2}\to 0$.
  This regime is weakly nonlinear only in the local but not (necessarily)
  in the 
  global sense and the intensity is allowed to have moderately large
  values. 
  As in the globally weak short wavelength regime
  the leading effect 
  is the shift of the  nonlinear wave number $\kappa_r$ (and 
  $\kappa_\eta$) which leads to phase shifts of order
  $\frac{g\ell  \overline{|\phi|^2}}{k}$. 
  As these phase shifts may be large we may \emph{not} expand the oscillatory 
  terms and the nonlinear effect in the wavefunction comes in the leading
  order. If we are only interested in the leading effect we may 
  neglect all other deformations altogether. 
  In this regime the equations that describe
  the stationary states on nonlinear quantum graphs simplify considerably
  in form but remain nonlinear.\\
  The explicit leading order of the wavenumber shift is consistent 
  as long as the intensity is only growing moderately
  as $\overline{|\phi|^2} =O(k)$ (at fixed $g$ and $\ell$).
  The regime, however, allows a larger growth
  $\overline{|\phi_e|^2} =o(k^2)$ but this requires to 
  calculate the nonlinear wavenumber
  $\kappa_r$ (and 
  $\kappa_\eta$ for general initial conditions) to all orders
  which is done in Appendix~\ref{angularfrequencies}.
\end{itemize}
In \cite{GW2}
we will consider a number of simple graph structures
as case studies how these regimes can be explored with our approach.
To come back to the discussion we had at the beginning of this section,
let us compare again our approach to any perturbation theory based on writing 
the wave function in the form $\phi(x)=\phi_0(x) + \delta\phi(x)$
where $\phi_0(x)$ is a solution of the corresponding linear equation
and $\delta\phi(x)$ a small perturbation.
In the asymptotic regimes R1 and R2 consistency requires that
equations \eqref{eq:real_perturbation_sol} are expanded further with respect
to small parameters which leads to the form $\phi(x)=\phi_0(x) + \delta\phi(x)$.
So our approach contains standard perturbation theory as a special case.
As regimes R1 and R2 can be obtained by linearization of the
of the NLSE with respect to $\delta \phi(x)$ no genuine nonlinear effects
such as bifurcations or multistability can be described.
Regime R3 however is not consistent with a small perturbation of the
wavefunction and cannot be obtained by linearization of the NLSE in 
standard perturbation theory. We will show that genuine nonlinear effects
can be described in this regime for sufficiently simple graph structures
in \cite{GW2}.
\\
If necessary it is not conceptually difficult
to obtain higher order approximations in the canonical perturbation
theory though the expressions become more and more cumbersome; using
symbolic computer algebra software is the obvious choice.

\section{Conclusion}
\label{conclusion}

In this paper we considered the stationary NLSE
on open and closed metric graphs, in short nonlinear
quantum graphs, as a model that makes it possible to investigate 
topological effects on nonlinear waves.  The solutions consist of stationary solutions of
the one-dimensional NLSE on the edges
(finite intervals or half lines) that obey matching conditions at the
edges.  We have given a complete qualitative description of all local
solutions, including solutions that may form singularities when
extended beyond the interval that represents a finite edge (bond).
Our qualitative analysis uses an exact equivalence between the
NLSE in one dimension and the dynamics of
a particle in two dimensions and a central potential where the spatial
variable $x$ takes the role of the time.  The chemical potential (or
energy) $\mu$, nonlinear coupling constant $g$, and the power of the
nonlinearity $\nu$ characterize both the 
NLSE and the corresponding central potential in the equivalent
dynamics. The Hamiltonian energy and the angular momentum are two
constants of motion for the equivalent dynamical system. They characterise the 
trajectories and hence the local
solutions of the NLSE.  Scaling properties
make it possible to reduce the analysis to $\mu=\pm 1$ and $g=\pm 1$.  For
$\nu=1$ we compile the complete set of analytic solutions 
in the Appendix; so far these have
been given explicitly only for solutions that remain bounded on an infinite
interval.

The knowledge of the solutions along each edge formally reduces the
problem of characterising the solutions on a graph to a finite set of
nonlinear equations that follow from the matching conditions and a
nonlinear transfer operator that expresses the wave function and its
derivative at one end of an edge in terms of the values at the other
end.  While these equations may be solved numerically complete
analytical solutions will generally be hard to find even for quite
simple graphs.  In order to simplify the nonlinear transfer operator
we have introduced a canonical perturbation theory for the 
NLSE valid for small $g$ (and arbitrary $\nu$). This
is a very powerful tool. In contrast to diagrammatic approaches 
which only 
yield corrections linear in $g$ for quantities such as the wave function,  
in the canonical perturbation theory $g$ enters the wave function in a
nonlinear way.

We have here focused on the NLSE on quantum graphs. Generalizations
to other nonlinear wave equation can be worked out in an analogous
way. Such generalizations may be necessary for physical applications
of the framework. For example, the NLSE appears in nonlinear optics in a
specific approximation where the envelope of an optical field is considered.
Backscattering from the vertices implies that this approximation may break
down and more general wave equations (not necessarily for a scalar wave)
need to be considered. For the qualitative understanding
of the combined effect of
nonlinear wave propagation and network topology the nonlinear quantum graphs 
based on the simpler NLSE is rich in complexity 
and may give already a lot of insight.

In the second paper of this series \cite{GW2} we will analyze some
basic closed and open graph structures analytically and numerically.
Among other things we will show that using the canonical perturbation theory
described here allows for an analytical description of
genuine non-linear effects.  We will also give an outlook on open questions.

\appendix

\section{Analytic Local Solutions on a Given Edge for $\nu=1$ in
  Terms of Jacobi Elliptic Functions}
\label{sec:analytic_solutions}

The bounded stationary solutions for the cubic NLSE ($\nu=1$) 
on the infinite line or a ring are
known and can be expressed in terms of elliptic functions
\cite{NIST,Gradsteyn}.  The construction of stationary solutions on
graphs requires the knowledge of \emph{all} local solutions including
those that are unbounded when continued to the infinite line.  These
can also be reduced to elliptic functions. As we are not aware that
these have been discussed in the literature we here give a complete
overview of all local solutions of the one-dimensional cubic stationary NLSE.  
Due to the scaling laws
\eqref{eq:scal1} and \eqref{eq:scal2} it is sufficient to consider
solutions
$$\phi(x)=R_{\pm 1, \pm 1}(x;r_0,p_\eta,H,\sigma)
e^{i\vartheta_{\pm 1, \pm 1}(x;r_0,p_\eta,H,\sigma)}$$ for chemical
potential $\mu = \pm 1$ and nonlinear coupling $g=\pm 1$.  
We also state expressions for the
integrated intensity (scaled number of particles)
\begin{equation}
  N_{\pm 1,\pm 1}(x; r_0, p_\eta, H, \sigma):= \int_0^x |\phi(x')|^2 dx'
  = \int_0^x R_{\pm 1,\pm 1}(x'; r_0, p_\eta, H, \sigma)^2 dx'
\end{equation}
over an interval $[0,x]$.

\subsection{Elliptic integrals and elliptic
  functions}\label{sec:elliptic}

We use the following definitions for elliptic integrals (the Jacobi
form)
\begin{subequations}
  \begin{align}
    F(x|m):=&\int_0^x \frac{1}{\sqrt{1-u^2}\sqrt{1-m\, u^2}} du\\
    K(m):=&F(1|m)\\
    E(x|m):=& \int_0^x \frac{\sqrt{1-m\, u^2} }{\sqrt{1-u^2}} du
    \\
    \Pi(x|a,m):=&\int_0^x \frac{1}{\sqrt{1-u^2}\sqrt{1-m\, u^2}(1-a\,
      u^2)} du
  \end{align}
  \label{specfunc}
\end{subequations}
where $0\le x \le 1$, $m \le 1$ and $a \le 1$. Note that our
definition allows
$m$ and $a$ to be negative.\\
The notation in the literature is far from uniform.  
Our
choice seems the most concise for the present context and it is
usually straight forward to translate our definitions into the ones of
any standard reference on special functions. For instance, the
\textit{NIST Handbook of Mathematical Functions} \cite{NIST} defines
the three elliptical integrals $F(\phi,k)$, $E(\phi,k)$ and $\Pi(\phi,
\alpha,k)$ by setting $x=\sin(\phi)$,
$m=k^2$, and $a=\alpha^2$ in our definitions above.\\
Jacobi's Elliptic function $\mathrm{sn}(x,m)$, the elliptic sine, is
defined as the inverse of $F(u|m)$
\begin{equation}
  u=\mathrm{sn}(x,m) \qquad \Leftrightarrow \qquad x=F(u|m)\ .
\end{equation}
This defines $\mathrm{sn}(x,m)$ for $x \in [0,K(m)]$. This is extended
to a periodic function with period $4 K(m)$ by requiring
$\mathrm{sn}(K(m)+x,m)=\mathrm{sn}(K(m)-x,m)$,
$\mathrm{sn}(-x,m)=-\mathrm{sn}(x,m)$ and $\mathrm{sn}(x+ 4
K(m),m)=\mathrm{sn}(x,m)$.  So, $\mathrm{sn}(x,m)$ is an elliptic
generalization of $\sin(x)$.  The corresponding elliptic cosine
$\mathrm{cn}(x,m)$ is obtained by requiring that it is a continuous
function satisfying
\begin{equation}
  \mathrm{cn}^2(x,m)+\mathrm{sn}^2(x,m)=1
\end{equation}
such that $\mathrm{cn}(0,m)=1$. It is useful to also define the
non-negative function
\begin{equation}
  \mathrm{dn}(x,m):= \sqrt{1 - m\, \mathrm{sn}^2(x,m)}.
\end{equation}
At $m=0$ and $m=1$ the elliptic functions can be expressed as
\begin{subequations}
  \begin{align}
    \mathrm{sn}(x,0)=& \sin x, &  \mathrm{sn}(x,1)=& \tanh x,\\
    \mathrm{cn}(x,0)=& \cos x, &  \mathrm{cn}(x,1)=& \cosh^{-1} x,\\
    \mathrm{dn}(x,0)=& 1, & \mathrm{dn}(x,1)=& \cosh^{-1} x\ .
  \end{align}
\end{subequations}
Derivatives of elliptic functions can be expressed in terms of
elliptic functions
\begin{subequations}
  \begin{align}
    \frac{d}{dx}\mathrm{sn}(x,m)=& \mathrm{cn}(x,m)\mathrm{dn}(x,m),\\
    \frac{d}{dx}\mathrm{cn}(x,m)=& -\mathrm{sn}(x,m)\mathrm{dn}(x,m),\\
    \frac{d}{dx}\mathrm{dn}(x,m)=& -m\,
    \mathrm{sn}(x,m)\mathrm{cn}(x,m)\ .
  \end{align}
\end{subequations}
The first of these equations implies that $u=\mathrm{sn}(x,m)$ is a
solution of the first order ordinary differential equation
\begin{equation}
  \frac{d u}{dx}=\sqrt{1-u^2}\sqrt{1-m u^2}\ .
  \label{eq:sn_ODE}
\end{equation} 

\subsection{Repulsive case with positive chemical potential}

In the main text we have used the constants of motion $p_\eta$ and $H$ as
parameters for the formal solutions for arbitrary nonlinear exponent
$\nu$. 
For $\nu=1$ a different (equivalent) set of real parameters that we denote 
by $\rho_i$ ($i=1,2,3$) are more
useful.  For the repulsive case with positive
chemical potential ($g=1$ and $\mu=1$) they are implicitly defined
(given arbitrary real values for
$p_\eta$ and $H$) by
\begin{subequations}
  \begin{align}
    \rho_1+\rho_2+\rho_3=& 2\\
    \frac{1}{\rho_1}+\frac{1}{ \rho_2}+ \frac{1}{\rho_3}=&
    \frac{2 H}{p_\eta^2}\\
    \rho_1 \rho_2 \rho_3=& 2 p_\eta^2
  \end{align} \label{rhoi}
\end{subequations}
or, equivalently, through the identity
\begin{equation}
  P(R):=(R^2-\rho_1)(R^2-\rho_2)(R^2-\rho_3)=R^6-2R^4+4HR^2-2 p_\eta^2
  \label{eq:rhodef_polynom}
\end{equation}
of real polynomials in $R^2$. Note that the sign of $p_\eta$ does
not enter the definition of the parameters $\rho_i$.  The differential
equation for the amplitude $R_{1,1}(x)$ then reduces to $2 R^2
\left(\frac{dR}{dx}\right)^2=P(R)$ where the left-hand side is
non-negative.  This implies that the solutions will have amplitudes in
the intervals
where $P(R)>0$.\\
Note that \eqref{eq:rhodef_polynom} defines $P(R)$ as a real
polynomial of order three in $R^2$.  We thus expect that either all
three $\rho_i$ are real, or two $\rho_i$
are complex and one is real.\\
In the first case with three real $\rho_i$ we may order them as $0\le
\rho_1\le \rho_2\le\rho_3\le 2$ where the first and last inequalities
follow straight forwardly from Eq.\ (\ref{rhoi}). The motion is either
bounded with $\rho_1\le R_{1,1}(x)^2\le \rho_2$ or unbounded with
$R_{1,1}(x)^2>\rho_3$.  If $\rho_2=\rho_3$ then three solutions
coexist: a bounded dark soliton with $\rho_1\le R_{1,1}(x)^2 \le
\rho_2$, a constant amplitude solutions $R_{1,1}(x)^2= \rho_2\equiv
\rho_3$ and an unbounded solution with $R_{1,1}(x)^2\ge \rho_2\equiv
\rho_3$.
\\
In the complex case we may choose $\rho_3$ real and write $\rho_1=
\xi+i \chi$ and $\rho_2= \xi-i\chi$. The motion is unbounded with
$R_{1,1}(x)^2>\rho_3$.  Note, that in either case $r_0\equiv
R_{1,1}(0)$ has to be chosen consistently with the inequalities valid
for $R_{1,1}(x)$.

\subsubsection{Bounded solutions: $\rho_1 \le R_{1,1}(x)^2 \le
  \rho_2\le \rho_3$}

With the initial conditions $R(0)=r_0$ and $\vartheta(0)=0$ the
bounded solution with real $\rho_i$ is given by
\begin{subequations}
  \begin{align}
    R_{1,1}(x)= & \sqrt{\rho_1 +\left(\rho_2-\rho_1\right) u(x)^2}
    \label{eq:type1a}
    \\
    \vartheta_{1,1}(x)= &\frac{p_{\eta}}{\sigma \beta \rho_1}
    \left[2n\Pi \left(1|-a,m\right) + (-1)^n\Pi\left(u(x) |-a,m
      \right) - \Pi(u_0|-a,m)\right]
    \\
    N_{1,1}(x)=&\rho_3 x-\frac{\rho_3-\rho_1}{\sigma \beta}
    \left[ 2nE(1|m) +(-1)^nE(u(x)|m)-E(u_0|m) \right]\\
    u(x)=& \, \mathrm{sn}\left(y_0+\sigma \beta x,m\right)
  \end{align}
  \label{eq:type1}
\end{subequations}
where $m=\frac{\rho_2-\rho_1}{\rho_3-\rho_1}$,
$a=\frac{\rho_2-\rho_1}{\rho_1}$, $\beta=
\sqrt{\frac{\rho_3-\rho_1}{2}}$,
$u_0=\sqrt{\frac{r_0^2-\rho_1}{\rho_2-\rho_1}}$, $y_0=F\left(u_0 |
  m\right)$ and $n \in \mathbb{Z}$ such that $\left| \frac{y_0+\sigma
    \beta x}{K(m)} -2n \right| \le 1$.  The first line
\eqref{eq:type1a} is the substitution that reduces the ordinary
differential equation $2 R^2 \left(\frac{dR}{dx}\right)^2=P(R)$ for
$R_{1,1}(x)$ to \eqref{eq:sn_ODE} for $u(x)$.
\\
For $\rho_2=\rho_1$ this reduces to a constant amplitude solution
\begin{subequations}
  \begin{align}
    R_{1,1}(x;r_0,p_\eta,H,\sigma)=& \sqrt{\rho_1}
    \\
    \vartheta_{1,1}(x;r_0,p_\eta,H,\sigma)=&
    \frac{p_{\eta}}{\rho_1} x
    \\
    N_{1,1}(x; r_0, p_\eta, H, \sigma)=& \rho_1 x \ .
  \end{align}
\end{subequations}

\subsubsection{Unbounded solutions: $\rho_1 \le \rho_2\le \rho_3\le
  R_{1,1}(x)^2$}

The unbounded solution for real $\rho_i$ is given by
\begin{subequations}
  \begin{align}
    R_{1,1}(x;r_0,p_\eta,H,\sigma)=& \sqrt{\rho_2
      +\left(\rho_3-\rho_2\right) \frac{1}{1-u(x)^2}}
    \\
    \vartheta_{1,1}(x;r_0,p_\eta,H,\sigma)=&
    \frac{p_{\eta}}{\rho_2}x-
    \frac{p_\eta(\rho_3-\rho_2)}{\sigma \beta \rho_2 \rho_3} \left[
      2n\Pi\left(1 |a,m \right) + (-1)^n\Pi\left(u(x) |a,m \right) -
      \Pi(u_0|a,m) \right]
    \\
    N_{1,1}(x; r_0, p_\eta, H, \sigma)=&
    \begin{cases}
      \rho_3 x- \frac{\rho_3-\rho_1}{\sigma \beta } \left[
        E\left(u(x)|m\right)-E(u_0|m)\right]+
      \\
      \qquad \frac{\rho_3-\rho_1}{\sigma \beta } \left[
        u(x)\sqrt{\frac{1-mu(x)^2}{1-u(x)^2}}-
        u_0\sqrt{\frac{1-mu_0^2}{1-u_0^2}}
      \right] & \text {if $n=0$;}\\[0.4cm]
      \infty & \text{if $n \neq 0$.}
    \end{cases}
    \\
    u(x)=& \, \mathrm{sn}\left(y_0+\sigma \beta x,m\right)
  \end{align}
  \label{eq:type2}
\end{subequations}
where $m=\frac{\rho_2-\rho_1}{\rho_3-\rho_1}$,
$a=\frac{\rho_2}{\rho_3}$, $\beta= \sqrt{\frac{\rho_3-\rho_1}{2}}$,
$u_0=\sqrt{\frac{r_0^2-\rho_3}{r_0^2-\rho_2}}$, $y_0=F\left(u_0 |
  m\right)$ and $n \in\mathbb{Z}$ such that $ \left|\frac{y_0+\sigma
    \beta x}{K(m)} - 2n\right| \le 1$.
                       
\subsubsection{Special case: $\rho_1 < \rho_2=\rho_3$}

Three solutions coexist: a bounded dark soliton solution, a constant
amplitude solution, and an unbounded solution.  The dark soliton can
be obtained from setting $\rho_2=\rho_3$ in \eqref{eq:type1} which
then reduces to
\begin{subequations}
  \begin{align}
    R_{1,1}(x;r_0,p_\eta,H,\sigma)=& \sqrt{\rho_1
      +\left(\rho_2-\rho_1\right) u(x)^2}
    \\
    \vartheta_{1,1}(x;r_0,p_\eta,H,\sigma)=&
    \frac{p_{\eta}}{\rho_2}x + \frac{p_{\eta}\sqrt{2}}{\sigma
      \rho_2 \sqrt{\rho_1}} \left[\arctan\left(\sqrt{a}\, u(x)\right)
      -\arctan\left(\sqrt{a}\, u_0\right) \right]
    \\
    N_{1,1}(x; r_0, p_\eta, H, \sigma)=& \rho_2x -
    \frac{\rho_2-\rho_1}{\sigma \beta} \left[u(x)-u_0 \right]
    \\
    u(x)=& \, \mathrm{tanh}\left(y_0+\sigma \beta x\right)
  \end{align}
  \label{eq:type_1a}
\end{subequations}
where $a= \frac{\rho_2-\rho_1}{\rho_1}$,
$\beta=\sqrt{\frac{\rho_2-\rho_1}{2}}$,
$u_0=\sqrt{\frac{r_0^2-\rho_1}{\rho_2-\rho_1}} $ and
$y_0=\mathrm{arctanh} \left(u_0 \right)$.  The constant amplitude
solution is given by
\begin{subequations}
  \begin{align}
    R_{1,1}(x;r_0,p_\eta,H,\sigma)=& \sqrt{\rho_2}
    \\
    \vartheta_{1,1}(x;r_0,p_\eta,H,\sigma)=&
    \frac{p_{\eta}}{\rho_2}x
    \\
    N_{1,1}(x; r_0, p_\eta, H, \sigma)=& \rho_2x\ ,
  \end{align}
  \label{eq:type2a}
\end{subequations}
and the unbounded solution is
\begin{subequations}
  \begin{align}
    R_{1,1}(x;r_0,p_\eta,H,\sigma)=& \sqrt{\rho_1
      +\left(\rho_2-\rho_1\right) u(x)^{-2}}
    \\
    \vartheta_{1,1}(x;r_0,p_\eta,H,\sigma)=&
    \frac{p_\eta}{\rho_2}x+ \frac{p_\eta\sqrt{2}}{\sigma \rho_2
      \sqrt{\rho_1}} \left[ \arctan\left(\frac{u(x)}{\sqrt{a}}\right)
      -\arctan\left(\frac{u_0}{\sqrt{a}}\right) \right]
    \\
    N_{1,1}(x; r_0, p_\eta, H, \sigma)=&
    \begin{cases}
      \rho_2 x + \frac{\rho_2-\rho_1}{\sigma \beta} \left[
        \frac{1}{u(x)} - \frac{1}{u_0} \right]
      & \text{if $y_0-\beta \sigma x>0$;}\\
      \infty & \text{if $y_0-\beta \sigma x\le 0 $ }
    \end{cases}
    \\
    u(x)=& \, \mathrm{tanh}\left(y_0-\sigma \beta x\right)
  \end{align}
  \label{eq:type2b}
\end{subequations}
where $\beta= \sqrt{\frac{\rho_2-\rho_1}{2}}$,
$u_0=\sqrt{\frac{r_0^2-\rho_1}{\rho_2-\rho_1}} $ and
$y_0=\mathrm{arctanh} \left(u_0 \right)$ .  Equations
\eqref{eq:type2a} and \eqref{eq:type2b} can be obtained from
\eqref{eq:type2} by performing appropriate limits $\rho_2 \to \rho_3$.

\subsubsection{Unbounded solutions: $R_{1,1}(x)^2 \ge \rho_3 >1$,  
  $\rho_1=\xi+ i \chi= \rho_2^*$}
This is given by
\begin{subequations}
  \begin{align}
    R_{1,1}(x;r_0,p_\eta,H,\sigma)= & \sqrt{\rho_3 + \gamma
      \frac{u(x)^2 (1-m\, u(x)^2)}{ 1-u(x)^2}}
    \\
    \vartheta_{1,1}(x;r_0,p_\eta,H,\sigma)= &
    \frac{p_{\eta}}{\rho_3\beta \sigma (a-b)} \Big[ (1-b) \left[
      2n\Pi(1|b,m) + (-1)^n\Pi\left(u(x) |b,m\right)- \Pi(u_0|b,m)
    \right]
    -\nonumber\\
    & \qquad \qquad (1-a)\left[ 2n \Pi(1|a,m) + (-1)^n\Pi\left(u(x)
        |a,m \right) - \Pi(u_0|a,m) \right] \Big]
    \\
    N_{1,1}(x; r_0, p_\eta, H, \sigma)=&
    \begin{cases}
      (\rho_3+\gamma) x -\frac{2\gamma}{\sigma \beta}
      \left[E\left(u(x)|m\right)- E(u_0|m)\right]
      +\\
      \qquad \frac{\gamma}{\sigma\beta}\left[ u(x)\sqrt{\frac{1-m\,
            u(x)^2}{1-u(x)^2}} - u_0\sqrt{\frac{1-m\, u_0^2}{1-u_0^2}}
      \right]
      & \text{if $n=0$}\\[0.5cm]
      \infty & \text{if $n\neq 0 $}
    \end{cases}
    \\
    u(x)=& \, \mathrm{sn}\left(y_0+\sigma \beta x,m\right)
  \end{align}
  \label{eq:type3}
\end{subequations}
where $\gamma = \sqrt{(\rho_3-\xi)^2 + \chi^2}$, 
$ m=\frac{\gamma-\rho_3+\xi}{2\gamma}$, $\beta= \sqrt{\gamma/2}$, $u_0=
\sqrt{\frac{\gamma+r_0^2-\rho_3}{\gamma-\rho_3+\xi}+
  \sqrt{\left(\frac{\gamma+r_0^2-\rho_3}{
        \gamma-\rho_3+\xi}\right)^2
    -\frac{2(r_0^2-\rho_3)}{\gamma-\rho_3+\xi}}}$,
$y_0=F\left(u_0 | m\right)$, $a=
\frac{\rho_3-\gamma+\sqrt{\xi^2+\chi^2}}{2\rho_3}$, $b=
\frac{\rho_3-\gamma-\sqrt{\xi^2+\chi^2}}{2\rho_3} $, and $ n
\in\mathbb{Z}$ such that $\left|\frac{y_0+\sigma \beta x}{K(m)} -
  2n\right| \le 1$.

\subsection{Repulsive case with negative chemical potential}

For the solutions with $g=1$ and $\mu=-1$ the three parameters
$\rho_i$ ($i=1,2,3$) are defined by
\begin{subequations}
  \begin{align}
    \rho_1+\rho_2+\rho_3=& -2\\
    \frac{1}{\rho_1}+\frac{1}{ \rho_2}+ \frac{1}{\rho_3}=&
    \frac{2 H}{p_\eta^2}\\
    \rho_1 \rho_2 \rho_3=& 2 p_\eta^2
  \end{align}
\end{subequations}
or, equivalently
\begin{equation}
  P(R)=(R^2-\rho_1)(R^2-\rho_2)(R^2-\rho_3)=R^6+2R^4+4HR^2-2 p_\eta^2 \ .
  \label{eq:rhodef_polynom1}
\end{equation}
As in the previous case either all three $\rho_i$ are real, or two
$\rho_i$
are complex and one is real.\\
In the first case with three real $\rho_i$ we may order them as
$\rho_1\le \rho_2\le 0 \le\rho_3\le 2$ where the second and third
inequalities can be shown straight forwardly.  In the complex case we
may choose $\rho_3$ real and write $\rho_1= \xi+i \chi$ and $\rho_2=
\xi-i\chi$.  In both cases the motion is unbounded with
$R_{1,-1}(x)^2>\rho_3$.

\subsubsection{Unbounded solutions: $\rho_1\le \rho_2 \le 0 \le \rho_3
  \le R_{1,-1}(x)^2$}

These solutions obey the same formulas as the unbounded motion
\eqref{eq:type2} -- note however, that the parameters $\rho_i$ have
different restrictions.  Analogously, the special case
$\rho_2=\rho_3=0$ can be obtained from \eqref{eq:type2b}.

\subsubsection{Unbounded solutions: $R_{1,-1}(x)^2 \ge \rho_3 >1$,
  $\rho_1=\xi+ i \chi= \rho_2^*$}

These solutions obey the same formulas as the unbounded motion
\eqref{eq:type3}.

\subsection{Attractive case with positive chemical potential}

For the solutions with $g=-1$ and $\mu=1$ the three parameters
$\rho_i$ ($i=1,2,3$) are defined by
\begin{subequations}
  \begin{align}
    \rho_1+\rho_2+\rho_3=& -2\\
    \frac{1}{\rho_1}+\frac{1}{ \rho_2}+ \frac{1}{\rho_3}=&
    \frac{2 H}{p_\eta^2}> 0\\
    \rho_1 \rho_2 \rho_3=& -2 p_\eta^2
  \end{align}
\end{subequations}
or, equivalently,
\begin{equation}
  P(R)=(\rho_1-R^2)(\rho_2-R^2)(\rho_3-R^2)=-R^6-2R^4+4HR^2-2 p_\eta^2 \ .
  \label{eq:rhodef_polynom2}
\end{equation}
The right-hand side defines $P(R)$ as real polynomial in $R^2$, so
either all $\rho_i$ are real, or one is real and two are complex
conjugates. The latter case can however be excluded. Indeed the
differential equation for the amplitude $R_{-1,1}(x)$ is of the form
$2 R^2\left(\frac{dR}{dx}\right)^2= P(R) \ge 0$ which requires $P(R)$
to be positive. However, if $\rho_3=\rho_2^*$, then $\rho_1<0$ (because
$\rho_1\rho_2\rho_3=\rho_1 |\rho_2|^2= -2 p_\eta^2<0$) and
$P(R)=(\rho_1-R^2)|\rho_2 - R^2|^2<0$ for all real values of $R$.\\
We are left with the case that all $\rho_i$ are real and we may order
them as $\rho_1\le 0 \le \rho_2 \le\rho_3$, where the first and second
inequalities can be shown straight forwardly.  The solution is bounded
with $\rho_2\le R_{-1,1}(x)^2 \le \rho_3$ and given by
\begin{subequations}
  \begin{align}
    R_{-1,1}(x;r_0,p_\eta,H,\sigma)= & \sqrt{\rho_3
      -\left(\rho_3-\rho_2\right) \frac{1-u(x)^2}{1-m\,u(x)^2}}
    =\sqrt{\rho_1+\left(\rho_2-\rho_1\right) \frac{1}{1-m\, u(x)^2}}
    \\
    \vartheta_{-1,1}(x;r_0,p_\eta,H,\sigma)= &
    \frac{p_\eta}{\rho_1}x +\frac{p_\eta(\rho_1-\rho_2)}{\sigma
      \beta \rho_1 \rho_2}\left[ 2n\Pi(1|a,m) + (-1)^n\Pi\left(
        u(x)|a,m \right) -\Pi(u_0|a,m) \right]
    \\
    N_{-1,1}(x; r_0, p_\eta, H, \sigma)= & \rho_1
    x+\frac{\rho_2-\rho_1}{\sigma \beta} \left[ 2n \Pi(1|m,m) +
      (-1)^n\Pi\left(u(x)|m,m\right)-\Pi(u_0|m,m) \right]
    \\
    u(x)=& \, \mathrm{sn}\left(y_0+\sigma \beta x,m\right)
  \end{align}
  \label{eq:type6}
\end{subequations}  
where $m=\frac{\rho_3-\rho_2}{\rho_3-\rho_1}$, 
$a=\frac{\rho_1}{\rho_2}m$, 
$\beta= \sqrt{\frac{\rho_3-\rho_1}{2}}$,
$u_0=\sqrt{\frac{1}{m}\frac{r_0^2-\rho_2}{r_0^2-\rho_1}}$, 
$y_0=F\left(u_0 | m\right)$ 
and $n \in \mathbb{Z}$ such that 
$\left|
\frac{y_0+\sigma \beta x}{K(m)} -2n \right| \le 1$.

\subsection{Attractive case with negative
  chemical potential}

For the solutions with $g=-1$ and $\mu=-1$  the three parameters
$\rho_i$ ($i=1,2,3$) are defined by
\begin{subequations}
  \begin{align}
    \rho_1+\rho_2+\rho_3=& 2\\
    \frac{1}{\rho_1}+\frac{1}{ \rho_2}+ \frac{1}{\rho_3}=&  
    \frac{2 H}{p_\eta^2}\\
    \rho_1 \rho_2 \rho_3=& -2 p_\eta^2 
  \end{align}
\end{subequations}
or, equivalently
\begin{equation}
  P(R)=(\rho_1-R^2)(\rho_2-R^2)(\rho_3-R^2)=-R^6+2R^4+4HR^2-2 p_\eta^2 \ .
\label{eq:rhodef_polynom3}
\end{equation}
Analogously to the previous case all $\rho_i$ have to be real with
$\rho_1\le 0 \le
\rho_2 \le\rho_3\le 2$. 
The solution is bounded with $\rho_2\le R_{-1,-1}(x)^2 \le \rho_3$
and the formulas \eqref{eq:type6} remain valid. \\
The special case of the soliton $\rho_2=\rho_1=0$ ($p_\eta=0$ and $H=0$)
deserves some attention as the expression \eqref{eq:type6} formally
vanishes. The limit $\rho_2,\rho_1 \to 0$ at fixed $r_0$ is however
not trivial. This solution is given explicitly by
\begin{subequations}
  \begin{align}
    R_{-1,-1}(x;r_0,p_\eta,H,\sigma)=
    &
      \frac{\sqrt{2}}{\mathrm{cosh} 
        \left(y_0 -\sigma  x \right)}
    \\
    \vartheta_{-1,-1}(x;r_0,p_\eta,H,\sigma)=
    & 
      0
    \\
    N_{-1,-1}(x; r_0, p_\eta, H, \sigma)=
    &
      \frac{2}{\sigma} \left(\tanh(y_0)  
        -\tanh\left(y_0 -\sigma x \right)\right)
  \end{align}
  \label{eq:type7a}
\end{subequations}  
where $y_0=\mathrm{arccosh}\left( \sqrt{2}/r_0\right)$.

\section{Exact expressions for the angular frequencies for the cubic
  NLSE}
\label{angularfrequencies}

The two angular frequencies $\kappa_r$ and $\kappa_\eta$ in Eq.\
\eqref{eq:wavenumbers_exact} follow from $\partial I_r/\partial H$ and
$\partial I_r/\partial I_\eta$ with $I_r(H,I_\eta)$ defined in
Eq.\ \eqref{eq:radial_action}.  In the cubic case $\nu=1$ they are
given explicitly by
\begin{equation}\label{eqq1}
  \frac{\partial I_r}{\partial H}=\frac{1}{\pi}\int_{r_-}^{r_+}
  \frac{dr}{
    \sqrt{\left(4Hr^2-2I^2_\eta\mp2r^4\pm r^6\right)/(2r^2)}}
\end{equation}
and
\begin{equation}\label{eqq2}
  \frac{\partial I_r}{\partial I_\eta}=-\frac{I_{\eta}}{\pi}
  \int_{r_-}^{r_+}
  \frac{dr}{
    r^2\sqrt{\left(4Hr^2-2I^2_\eta\mp2r^4\pm r^6\right)/(2r^2)}}
\end{equation}
with $r_-$ and $r_+$ the two turning points of the dynamics that obey
the condition ${r_-}<{r_+}$.  We calculate them here explicitly
for $\mu=\pm 1$ and $g=\pm 1$. We include here the unbounded solutions
where $r_+\equiv \infty$.

\subsection{Repulsive case with positive chemical potential}
\subsubsection{Bounded solutions: $\rho_1\leq
  R_{1,1}(x)^2\leq\rho_2\leq\rho_3$}
Here the $r$-variable in Eqs.\ (\ref{eqq1}) and (\ref{eqq2}) is
restricted to values between $\sqrt{\rho_1}$ and
$\sqrt{\rho_2}$. Using the parametrization in Eq.\
(\ref{eq:rhodef_polynom}) we can express the integral in (\ref{eqq1})
as
\begin{equation}\label{eqq3}
  \frac{\partial I_r}{\partial H}=
  \frac{1}{\pi}\int_{\sqrt{\rho_1}}^{\sqrt{\rho_2}}
  \frac{dr}{\sqrt{\left(r^2-\rho_1\right)
      \left(r^2-\rho_2\right)\left(r^2-\rho_3\right)/(2r^2)}}
\end{equation}
and the one in (\ref{eqq2}) as
\begin{equation}\label{eqq4}
  \frac{\partial I_r}{\partial I_\eta}=
  -\frac{I_{\eta}}{\pi}\int_{\sqrt{\rho_1}}^{\sqrt{\rho_2}}
  \frac{dr}{r^2\sqrt{\left(r^2-\rho_1\right)
      \left(r^2-\rho_2\right)\left(r^2-\rho_3\right)/(2r^2)}}.
\end{equation}
The final result is
\begin{equation}
  \frac{\partial I_r}{\partial H}=\frac{1}{\pi\beta}K(m)
\end{equation}
and
\begin{equation}
  \frac{\partial I_r}{\partial I_\eta}
  =-\frac{I_{\eta}}{\pi\beta\rho_1}\Pi\left(1|a,m\right)
\end{equation}
with the functions $K(m)$ and $\Pi\left(1|a,m\right)$ defined in Eq.\
(\ref{specfunc}), $m$, $\beta$ and $a$ are defined in terms of
$\rho_1$, $\rho_2$ and $\rho_3$ after Eq.\ (\ref{eq:type1}).

\subsubsection{Unbounded solutions $\rho_1\leq\rho_2\leq\rho_3\leq
  R_{1,1}(x)^2$}\label{appB2}

In this case the $r$-dynamics is restricted by the condition
$r\geq\sqrt{\rho_3}$, we obtain
\begin{equation}
  \frac{\partial I_r}{\partial H}=\frac{1}{\pi\beta}K(m)
\end{equation}
and
\begin{equation}
  \frac{\partial I_r}{\partial I_\eta}=
  -\frac{I_\eta}{\pi}
  \left[\frac{1}{\rho_2\beta}K(m)-
    \frac{\rho_3-\rho_2}{\rho_2\rho_3\beta}
    \Pi\left(1|a,m\right)\right]
\end{equation}
with $m$, $\beta$ and $a$ as defined after Eq.\ (\ref{eq:type2}).

\subsubsection{Special case: $\rho_1<\rho_2=\rho=3$}

Due to the singularity resulting from the term proportional to
$\left(r^2-\rho_2\right)^{-1}$ the considered quantities tend to
infinity in this case.

\subsubsection{Unbounded solutions:
  $R_{1,1}\geq\rho_3>1,\,\rho_1=\xi+i\chi=\rho_2^*$}\label{appB4}

Here the motion takes place in the region $r\geq\sqrt{\rho_3}$, the
quantities of interest are given by
\begin{equation}
  \frac{\partial I_r}{\partial H}=\frac{1}{\pi\beta}K(m)
\end{equation} 
and
\begin{equation}
  \frac{\partial I_r}{\partial I_\eta}=-
  \frac{I_\eta}{\pi\beta(a-b)}
  \left[(1-b)\Pi(1|b,m)-(1-a)\Pi(1|a,m)\right]
\end{equation}
with $m$, $\beta$, $a$ and $b$ defined after Eq.\ (\ref{eq:type3}).

\subsection{Repulsive case with negative chemical potential}

Here unbounded motion is obtained, Eqs.\ (\ref{eqq3}) and (\ref{eqq4})
remain valid and the results from (\ref{appB2}) and (\ref{appB4}) remain
applicable.

\subsection{Attractive case with positive chemical potential}

Here Eqs.\ (\ref{eqq3}) and (\ref{eqq4}) are changed to
\begin{equation}\label{eqq5}
  \frac{\partial I_r}{\partial H}=
  \frac{1}{\pi}\int_{\sqrt{\rho_2}}^{\sqrt{\rho_3}}
  \frac{dr}{\sqrt{\left(\rho_1-r^2\right)\left(\rho_2-r^2\right)
      \left(\rho_3-r^2\right)/(2r^2)}}
\end{equation}
and
\begin{equation}\label{eqq6}
  \frac{\partial I_r}{\partial I_\eta}=
  -\frac{I_{\eta}}{\pi}\int_{\sqrt{\rho_2}}^{\sqrt{\rho_3}}
  \frac{dr}{r^2\sqrt{\left(\rho_1-r^2\right)\left(\rho_2-r^2\right)
      \left(\rho_3-r^2\right)/(2r^2)}}
\end{equation}
and finally result in
\begin{equation}
  \frac{\partial I_r}{\partial H}=\frac{1}{\pi\beta}K(m)
\end{equation}
and
\begin{equation}
  \frac{\partial I_r}{\partial I_\eta}=
  -\frac{I_\eta}{\rho_1\beta\pi}K(m)
  -\frac{I_\eta(\rho_1-\rho_2)}{\beta\rho_1\rho_2\pi}\Pi(1|a,m)
\end{equation}
with $\beta$, $m$ and $a$ defined after Eq.\ (\ref{eq:type6}).

\subsection{Attractive case with negative chemical potential}

In this case the expressions from the last section remain valid, in
the special case $\rho_2=\rho_1=0$, $\partial I_r/\partial H$ and
$\partial I_r/\partial I_\eta$ diverge due to the singularity of
the $r$-integral at zero.

\begin{acknowledgments}
  We would like
  to thank Uzy Smilansky for the support he gave to this work and 
  for sharing his thoughts and ideas during research
  stays of both authors at the Weizmann Institute. 
  We would like to thank the Weizmann Institute for its hospitality.
  D.W. acknowledges the Minerva foundation for financial support making 
  this research stay possible.
\end{acknowledgments}


\begin{thebibliography}{10}
\bibitem{footnote1} Any choice of a coordinate system as
described above implicitly implies a choice of direction on each edge
-- however we treat all coordinate systems that attribute the same
distance to any two points as equivalent. For, example, on each bond we may
switch from the initial coordinates to the equivalent coordinates
$\hat{x}_{b}= \ell_{b}- x_b$ which traverse the bond in the opposite
direction.
\bibitem{footnote2} 
  Vertices of valency two with standard matching conditions
  and the same value of the nonlinear coupling constant $g$ 
  on either side can
  effectively be deleted by joining the two adjacent edges to one
  single edge of combined length.
\end{thebibliography}
\end{document}